\documentclass[bibyear]{aa}  
\usepackage{longtable} 
\usepackage{pdflscape} 
\usepackage{multicol}
\usepackage{supertabular,booktabs}
\usepackage[normalem]{ulem}
\usepackage{amsmath}

\usepackage{placeins}
\usepackage{graphicx}
\usepackage[flushleft]{threeparttable}
\usepackage{txfonts}
\usepackage{diagbox}
\makeatletter
\renewcommand*\aa@pageof{, page \thepage{} of \pageref*{LastPage}}
\makeatother
\usepackage{natbib,twoopt}
\usepackage[breaklinks=true,linkcolor=blue,allcolors=blue,colorlinks=true]{hyperref} 
\bibpunct{(}{)}{;}{a}{}{,}             
\makeatletter
  \newcommandtwoopt{\citeads}[3][][]{\href{http://adsabs.harvard.edu/abs/#3}%
    {\def\hyper@linkstart##1##2{}%
     \let\hyper@linkend\@empty\citealp[#1][#2]{#3}}}
  \newcommandtwoopt{\citepads}[3][][]{\href{http://adsabs.harvard.edu/abs/#3}%
    {\def\hyper@linkstart##1##2{}%
     \let\hyper@linkend\@empty\citep[#1][#2]{#3}}}
  \newcommandtwoopt{\citetads}[3][][]{\href{http://adsabs.harvard.edu/abs/#3}%
    {\def\hyper@linkstart##1##2{}%
     \let\hyper@linkend\@empty\citet[#1][#2]{#3}}}
  \newcommandtwoopt{\citeyearads}[3][][]%
    {\href{http://adsabs.harvard.edu/abs/#3}
    {\def\hyper@linkstart##1##2{}%
     \let\hyper@linkend\@empty\citeyear[#1][#2]{#3}}}
\makeatother
\usepackage{orcidlink}

\usepackage{booktabs}
\usepackage{colortbl}

\begin{document} 

    \title{Gas excitation in galaxies and active galactic nuclei with \protect{$\rm \ion{He}{II} \lambda 4686$} and X-ray emission}
   \titlerunning{X-rays and $\rm \ion{He}{II} \lambda 4686$ in galaxies and AGN}

   \author{K. Kouroumpatzakis
          \inst{1}\,\orcidlink{0000-0002-1444-2016}\fnmsep\thanks{email:konstantinos.kouroumpatzakis@asu.cas.cz}
          \and 
          J. Svoboda\inst{1}\,\orcidlink{0000-0003-2931-0742}
          }

   \institute{
            Astronomical Institute, Academy of Sciences, Boční II 1401, CZ-14131 Prague, Czech Republic\\ 
             }

   \date{Received november 27, 2024; accepted March 3, 2025}

\defcitealias{2012MNRAS.421.1043S}{SB12}

 
  \abstract
   {
   The origin of $\rm \ion{He}{II}$ emission in galaxies remains a debated topic, requiring ionizing photons with energies exceeding 54~eV. 
   While massive stars, such as Wolf-Rayet stars, have been considered potential sources, their UV flux often fails to fully explain the observed \ion{He}{II} emission. 
   Recent studies suggest that X-ray binaries (XRBs) might contribute significantly to this ionization.
   }
   {
   We explore the relationship between X-ray and $\rm \ion{He}{II} \lambda 4686$ emission in a statistically significant sample of galaxies, investigating whether X-ray sources, including active galactic nuclei (AGNs) and XRBs, serve as the primary mechanism for \ion{He}{II} ionization across different galactic environments.
   }
   {
    We cross-matched a sample of known well-detected \ion{He}{II} galaxies with the Chandra Source Catalog, yielding 165 galaxies with X-ray and $\rm \ion{He}{II} \lambda 4686$ detections. 
    The sources were classified into star-forming galaxies (SFGs) and AGNs based on the BPT diagram and a classification scheme defined for \ion{He}{II} galaxies. 
    We analyzed the correlation between X-ray and \ion{He}{II} luminosity across different energy bands and other parameters.
   }
   {
   The comparison between X-ray and \ion{He}{II} luminosity shows a strong, linear correlation across AGNs and SFGs spanning over seven orders of magnitude. 
   AGNs generally exhibit higher \ion{He}{II}/\ion{H}{$\beta$} flux ratios, stronger extinction, and harder X-ray spectra.
   The O32 ratio of SFGs is tightly correlated with the \ion{H}{$\beta$} equivalent width ($\rm EW_{H\beta}$) but not with the \ion{He}{II}/\ion{H}{$\beta$} ratio, suggesting a different excitation mechanism.
   We derive an O32--$\rm EW_{H\beta}$ line above which only AGNs of our sample reside.
   }
   {
   The tight correlation between X-ray and \ion{He}{II} luminosity supports X-rays as the primary driver of \ion{He}{II} excitation. 
   While AGNs have one common ionization source, the central black hole, in SFGs low-energy species are mainly excited by UV emission related to star-forming activity, however, high-energy species like \ion{He}{II} require the presence of XRBs. 
   }


   \maketitle
%
\section{Introduction}
\label{sec:intro}

Doubly ionized Helium with emission in the wavelengths $\lambda =$ 1640, 2733, 3203, 4686, 5412, and 8237~\AA\ has a very high excitation threshold of 54~eV.
Thus, it requires the presence of hard ionizing radiation with wavelengths $\lambda \leq 228$~\AA.
It has long been debated which is the excitation mechanism responsible for the \ion{He}{II} emission.
Young massive stars are a source of ionizing radiation, but their UV emission energy has been considered not high enough to excite \ion{He}{II} \citep[e.g.,][]{2015A&A...576A..83S,2019A&A...624A..89N,2019A&A...621A.105S,2019MNRAS.490..978P,2020A&A...636A..47S,2021ApJ...922..170B,2022ApJ...938...16O}.
Modeling the radiation of simple stellar populations (SPs) showed that their flux at the energies of 54~eV can be down to about four orders of magnitude lower than that required to ionize hydrogen or oxygen.

Wolf-Rayet (WR) stars were known to show  \ion{He}{II} emission in their spectra \citep[e.g.,][]{1964SvA.....8..172R,1989A&A...210..236S}, so they were naturally considered as the underlying source of the \ion{He}{II} emission found in the spectra of galaxies \citep[e.g.,][]{1996ApJ...467L..17S}.
The WR stars can be mainly found in galaxies with developed SPs and metallicities $\rm 12+log(O/H) \ge 8.4$ but also actively star-forming \citep[e.g.,][]{1980A&A....90L..17M}.
However, on most occasions, WR stars' emission could not fully account for the observed fluxes \citep[e.g.,][]{2012MNRAS.421.1043S,2013ApJ...766...91J}.

The observed $\rm \ion{He}{II} \lambda 4686$ intensity was found to show an anti-correlation with metallicity and SP age \citep[e.g.,][]{1986Natur.322..511P,2019A&A...622L..10S}, similar to that of the X-ray radiation from X-ray binary (XRB) populations \citep[e.g.,][]{2013ApJ...776L..31F,2016MNRAS.457.4081B}.
Led from such similarities some studies suggested XRBs or ultraluminous XRBs (ULXs) as the main sources of its ionization \citep[e.g.,][]{1991ApJ...373..458G,2019A&A...622L..10S,2019A&A...627A..63O,2021A&A...656A.127S,2022A&A...661A..67O,2022ApJ...930..135L,2024arXiv240508121T}.
Modeling the ionizing capabilities of simple stellar and their XRB populations showed that the presence of the latter extends the high-energy ionizing radiation which affects the strength of nebular lines like $\rm \ion{He}{II}$ \citep{2019A&A...622L..10S,2022ApJ...930...37U,2024ApJ...960...13G}.

On the other hand, the study of \cite{2020MNRAS.496.3796S} showed no significant difference in the X-ray emission of galaxies with or without $\rm \ion{He}{II} \lambda 1640$ ionization suggesting that the former cannot be the main excitation mechanism of the latter, albeit the small size of the studied sample.
Other alternative sources of \ion{He}{II} excitation involve galaxies with extended initial mass function (IMF) hosting supermassive stars with stellar mass $M_\star \gtrsim 100 ~M_\odot$ or the presence of metal-free population III stars \citep[e.g.,][]{2013A&A...556A..68C,2015A&A...581A..15S,2016MNRAS.458..624C,2019A&A...629A.134G}.
This solution was also suggested by recent observations with the James Webb Space Telescope to explain the $\rm \ion{He}{II} \lambda 4686$ emission of a distant $z=11$ galaxy \citep[][]{2024A&A...687A..67M}.

In the modern Universe, \ion{He}{II} emission is a strikingly scarce phenomenon among galaxies. 
Analyzing the Sloan Digital Sky Survey (SDSS) Data Release (DR) 7, which encompasses the spectra of approximately one million galaxies and quasars, \cite{2012MNRAS.421.1043S} (hereafter \citetalias{2012MNRAS.421.1043S}) identified only 2,865 well-detected sources with signal-to-noise ratio (S/N) > 5.5 in $\rm \ion{He}{II} \lambda 4686$ emission. 
This rarity highlights the exceptional astrophysical conditions required for \ion{He}{II} excitation and underscores the critical need to investigate its underlying mechanisms and contributing factors.

To explore the effect X-ray emission has on exciting \ion{He}{II}\footnote{from now on and throughout this paper \ion{He}{II} refers to $\ion{He}{II} \lambda 4686$}, a dedicated study based on a statistically significant sample of galaxies is required.
A potential correlation could reveal an underlying cause or on the other hand, lack of it would direct us to some other potential mechanism that can excite \ion{He}{II}.
This work combines the \citetalias{2012MNRAS.421.1043S} sample of \ion{He}{II} emission galaxies with the Chandra source catalog \citep[CSC;][]{2010ApJS..189...37E,2024ApJS..274...22E} to study the observational correlation between the X-ray and \ion{He}{II} emission.

This paper is organized as follows: Section \ref{sec:sample} describes the X-ray and \ion{He}{II} sample.
Section \ref{sec:results} provides details about the classification of the sources, describes the main analysis, and gives the results.
Section \ref{sec:All_HeII} expands the analysis by examining some of the ionization properties of all the \citetalias{2012MNRAS.421.1043S} galaxies regardless if they were observed in X-rays.
Finally in Section \ref{sec:discussion} we discuss based on the results and in Section \ref{sec:conclusions} we summarize this work.
Throughout this paper we adopted a \cite{2016A&A...594A..13P} cosmology ($\rm \Omega_m=0.308$, $h=0.678$).
Regarding the statistical analyses, confidence intervals (CIs) and errors refer to 32\%68\% and Scott's rule \citep{10.1093/biomet/66.3.605} is used to determine the size of the histogram's bins.

\section{Sample of \texorpdfstring{\ion{He}{II}}{} and X-ray galaxies}
\label{sec:sample}

To study the X-ray emission of \ion{He}{II} emission galaxies we adopt the sample of \citetalias{2012MNRAS.421.1043S}.
It provides the \ion{He}{II}/\ion{H}{$\beta$} and [\ion{N}{II}]/\ion{H}{$\alpha$} flux ratios for 2,865 galaxies that have detected $\rm \ion{He}{II} $ with S/N~$> 5.5$.
We crossmatch the \citetalias{2012MNRAS.421.1043S} sample with the MPA-JHU catalog \citep{2003MNRAS.346.1055K,10.1111/j.1365-2966.2004.07881.x,Tremonti_2004} which provides the fluxes of several emission lines and their uncertainties.
We also adopt the MPA-JHU estimations of the galaxies' star-formation rate (SFR), stellar mass ($M_\star$), and metallicity (12~+~log(O/H)).

Moreover, we crossmatch with CSC, providing uniformly calibrated and analyzed X-ray fluxes in the energy bands 0.5--1.2 (soft), 1.2--2 (medium), 2-7 (hard), and 0.5--7~keV (total). 
CSC accumulates the detected sources of the Chandra Space Telescope (Chandra from now on) public archive.
The estimated fluxes are based on a Bayesian X-ray aperture photometry code.
The second release of CSC has a compact source sensitivity limit of about five photons across most of Chandra's field of view. 
The cross-matching process led to 254 galaxies. 
After omitting sources that had bad X-ray photometries indicated by CSC's \texttt{confidence}, \texttt{saturation}, or \texttt{readout streak} flags or problematic (e.g., infinite) fluxes the final sample is composed of 165 bona fide $\rm \ion{He}{II}$ and X-ray emission galaxies.
The luminosities are calculated based on the fluxes and the distances estimated through the redshifts provided by MPA-JHU.
The classification of the sources is examined in Section \ref{sec:classification}.

The SFRs used in this work are adopted by the MPA-JHU analysis.
They were calculated with the methods of \cite{2005MNRAS.362...41G} and \cite{2007ApJS..173..267S} which adopted a \cite{2003PASP..115..763C} IMF.
To homogenize our comparisons we transfer all relations used in the following analysis to the \cite{2003PASP..115..763C} IMF used by our sample's SFR estimations.
Thus, we multiplied the relations of \cite{2014MNRAS.437.1698M} and \cite{2016MNRAS.457.4081B} by 0.67 (following \cite{2014ARA&A..52..415M}) to convert from \cite{1955ApJ...121..161S} IMF. 
Similarly, we multiply by 1.06 to convert the \cite{2013ApJ...776L..31F} and \cite{2021ApJ...907...17L} relations which had adopted the \cite{2001MNRAS.322..231K} IMF.
Moreover, the CSC provides fluxes in the energy range of 0.5-7~keV while most relations were established on the 0.5-8~keV energy bands.
Assuming a typical power-law spectrum with $N_{\rm H} = 10^{21}~{\rm cm}^{-2}$, and $\Gamma = 2$ we estimate a conversion of 0.95 which equals to $-0.02$~dex to account for the shorter energy range.
Similarly, the conversion from 0.3--8~kev to 0.5-7~keV is equal to 0.89 or $-0.05$~dex.

\section{Results}
\label{sec:results}

\begin{figure*}[ht!]
    \centering
    \includegraphics[width=1\textwidth]{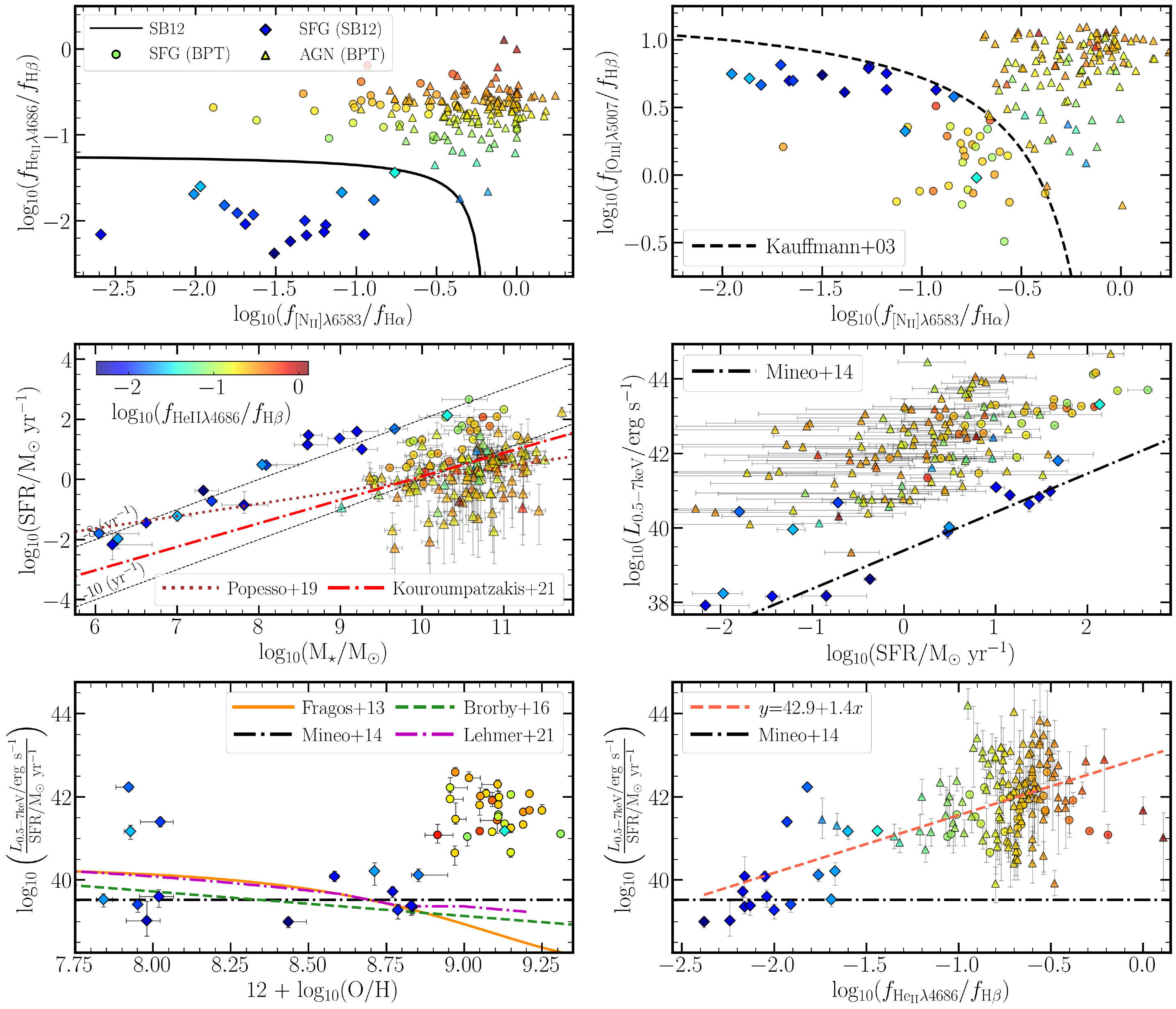}
    \caption{Physical relations of galaxies emitting \ion{He}{II} with $\rm SnR>5.5$, measured and cataloged by \citetalias{2012MNRAS.421.1043S}, detected in X-rays and cataloged in the CSCv2.0.
    In all subplots, the points are color-coded based on the galaxies' \ion{He}{II}/\ion{H}{$\beta$} ratio.
    The emission-line ratios have low uncertainties which are not plotted for clarity.
    BPT SFGs are represented by a circle, and BPT AGNs are represented by a triangle marker. 
    \citetalias{2012MNRAS.421.1043S} SFGs are marked with a rhombus. 
    Top left: \ion{He}{II}/\ion{H}{$\beta$} ratio as a function of the [\ion{N}{II}]/\ion{H}{$\alpha$} ratio. 
    The black curve represents the \citetalias{2012MNRAS.421.1043S} classification curve separating AGNs (top right) from SFGs (bottom left).
    Top right: [\ion{O}{III}]/\ion{H}{$\beta$} as a function of the [\ion{N}{II}]/\ion{H}{$\alpha$} ratio, aka the BPT diagram. 
    The dashed black line shows the \cite{2003MNRAS.346.1055K} classification curve above which composite galaxies and AGNs are located.
    Middle left: SFR as a function of $M_\star$, aka the main sequence of SFGs.
    The brown-dotted, and red dashed-dotted lines represent the
    \protect\cite{2019MNRAS.483.3213P}, and \protect\cite{2021MNRAS.506.3079K} main sequence fits respectively.
    Middle right: $L_{\rm X}$ as a function of SFR.
    The black dashed-dotted line represents the \cite{2014MNRAS.437.1698M} relation.
    Bottom left: $L_{\rm X}$/SFR as a function of metallicity.
    The orange continuous, black dashed-doted, green dashed, and magenta dashed-dotted lines represent the \cite{2013ApJ...776L..31F}, \cite{2014MNRAS.437.1698M}, \cite{2016MNRAS.457.4081B}, and \cite{2021ApJ...907...17L} relations respectively.
    Bottom right: $L_{\rm X}$/SFR as a function of \ion{He}{II}/\ion{H}{$\beta$} ratio. 
    The black dashed-dotted line represents the \cite{2014MNRAS.437.1698M} relation.
    The red dashed line shows the best linear regression fit involving all sources regardless of their class.
    }
    \label{fig:Lx_HeII_All}
\end{figure*}

Several comparisons presented in Figure \ref{fig:Lx_HeII_All} allow us to have a complete overview of the properties of the selected sample.
The top two plots show the classifications based on BPT  \citep{1981PASP...93....5B,1987ApJS...63..295V,2001ApJ...556..121K,2003MNRAS.346.1055K} and \citetalias{2012MNRAS.421.1043S} diagrams. 
The middle left plot presents the SFR over $M_\star$ of the galaxies, known as the main sequence of star-forming galaxies (SFGs), and the middle right plot the relation between their X-ray luminosity ($L_{\rm X}$) and SFR.
The two lower plots present the normalized-by-SFR X-ray luminosity ($L_{\rm X}$/SFR) of the sources as a function of metallicity (lower left) and $\rm \ion{He}{II}$/\ion{H}{$\beta$} ratio (lower right).
In all plots of Figure \ref{fig:Lx_HeII_All}, all points are color-coded based on the sources' \ion{He}{II}/\ion{H}{$\beta$} ratio.

\subsection{Classification of the \texorpdfstring{\ion{He}{II}}{} and X-ray galaxies}
\label{sec:classification}

The top two plots of Figure \ref{fig:Lx_HeII_All} show the \ion{He}{II}/\ion{H}{$\beta$} over [\ion{N}{II}]/\ion{H}{$\alpha$} flux ratio diagram, and the BPT [\ion{O}{III}]/\ion{H}{$\beta$} over [\ion{N}{II}]/\ion{H}{$\alpha$} diagram. 
Based on the BPT diagram, the \cite{2001ApJ...556..121K} theoretical extreme ionization line, and the \cite{2003MNRAS.346.1055K} empirical line, 46 galaxies are classified as star-forming and the rest are active galactic nuclei (AGNs) or composite galaxies.
Based on the classification proposed by \citetalias{2012MNRAS.421.1043S} only 17 sources are classified as SFGs. 
It must be noted that many BPT SFGs are not considered as SFGs by the \citetalias{2012MNRAS.421.1043S} diagram because they have high \ion{He}{II}/\ion{H}{$\beta$} ratios although they show relatively low \ion{N}{II}/\ion{H}{$\alpha$} ratios (Figure \ref{fig:Lx_HeII_All} upper left).
These sources are mostly located in the lower ionization parameter (U) and higher metallicity part of the BPT \citep[e.g.,][]{2001ApJ...556..121K}. 

\begin{figure*}[ht!]
    \centering
    \includegraphics[width=0.9\textwidth]{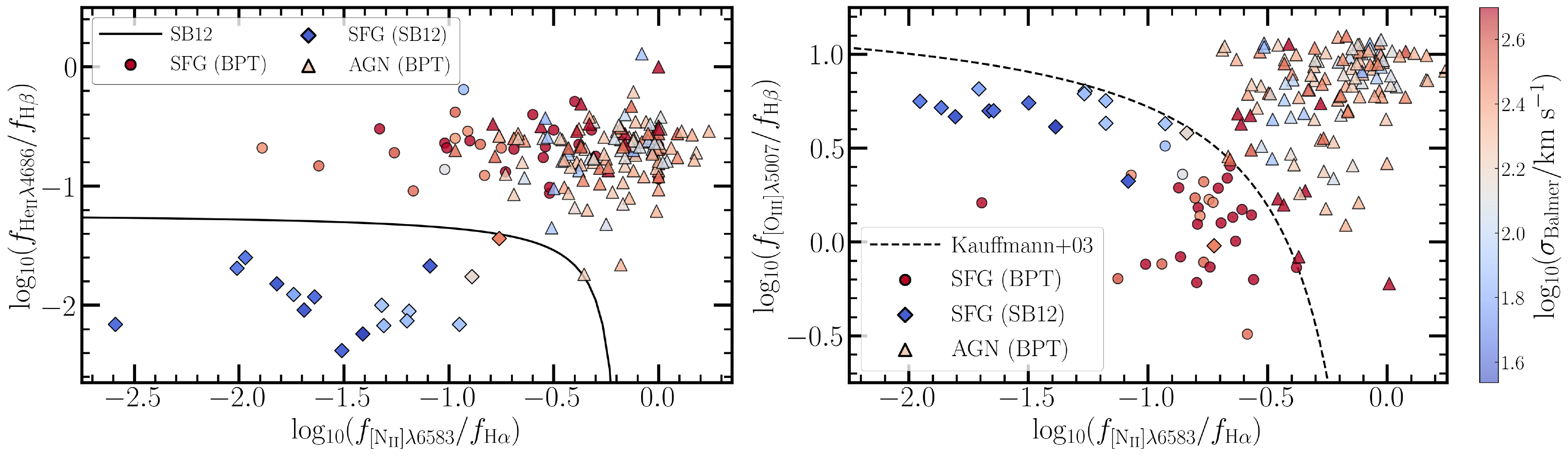}
    \includegraphics[width=0.9\textwidth]{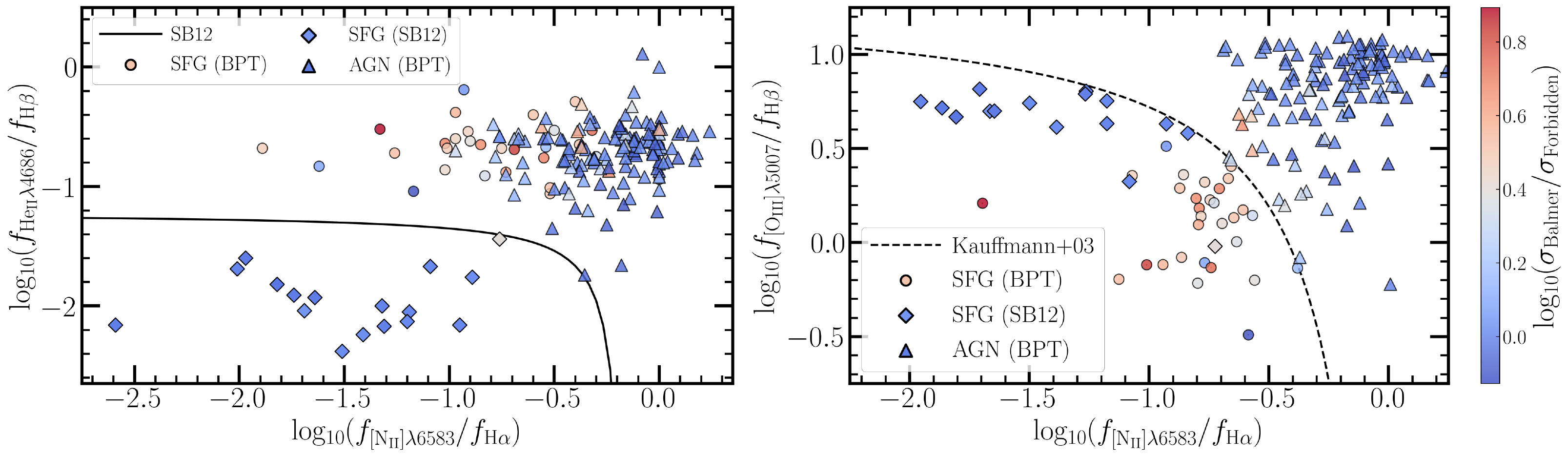}
    \caption{
    Same plots as the upper plots of Figure \ref{fig:Lx_HeII_All} but color-coded based on the galaxies' velocity dispersion ($\sigma$) of their Balmer lines (top), and the ratio of the velocity dispersions of the Balmer and forbidden lines (bottom). 
    }
    \label{fig:BPT_SB12_SIGMA}
\end{figure*}

Moreover, as it is shown in the $L_{\rm X}/{\rm SFR}$--metallicity plot of Figure \ref{fig:Lx_HeII_All} (lower left) these sources form a separate group of high metallicity, therefore developed galaxies. 
This group also shows a large excess of X-ray emission, which indicates the presence of AGN.
We further explore the classification of the galaxies by examining the velocity dispersion ($\sigma$) of their Balmer lines, as measured by the MPA-JHU catalog (Figure \ref{fig:BPT_SB12_SIGMA}).
The group of BPT-SFGs-but-SB12-AGNs shows high velocity dispersions, strongly suggesting the presence of AGN. 
Interestingly, their velocity dispersion is even higher than the BPT AGNs.

In a simplified archetypical scenario, AGNs are divided into Seyfert I (aka quasars or broad-line AGNs), and Seyfert II \citep[aka narrow-line AGNs; e.g.,][]{1974ApJ...192..581K,1993ARA&A..31..473A}.
For the former, we can observe the AGN's broad-line region (BLR) because we see them face-on in our line of sight. 
Due to edge-on inclination, we can not observe the BLR of Seyfert II AGNs, which are otherwise located at the top right of the BPT diagram.
AGNs showing broadened Balmer lines are not usually included in the BPT diagram due to difficulties in disentangling the narrow from the broad component \citep[e.g.,][]{1987ApJS...63..295V,2003MNRAS.346.1055K}.
BLRs are characterized by high-density gas, typically exceeding $n_e > 10^9 ~{\rm cm^{-3}}$.
These higher densities lead to collisions dominating the de-excitation of the gas and the suppression of the forbidden emission lines.
Therefore, a significantly higher Balmer-to-forbidden line velocity dispersion ratio can help us distinguish whether the high-$\sigma$ is attributed to the BLR of an AGN or is driven by shocks related to extreme star-forming activity (e.g., recent supernovae or \ion{H}{II} bubbles).

In the lower plots of Figure \ref{fig:BPT_SB12_SIGMA} we examine the possibility of this high $\sigma$ being a result of a shock-dominated interstellar medium.
The sources are color-coded based on the ratio of their Balmer-to-forbidden velocity dispersion.
Most BPT-SFGs-but-SB12-AGNs show a significantly higher ratio while their $\sigma$ is overall higher.
Additionally, we present in Appendix \ref{sec:Appendix_A} the optical spectra near the \ion{H}{$\alpha$} and \ion{H}{$\beta$} lines of this sample's X-ray detected BPT-SFGs-but-SB12-AGNs.
Based on visual inspection, 29/30 sources clearly show wider Balmer lines signaling the contribution from a BLR \citep[e.g.,][]{2013ApJ...775..116R,2015ApJ...813...82R}. 
Because these sources have broadened Balmer lines but relatively lower $\sigma$ compared to those defined as broad-line AGNs ($\sigma > 2000 ~ {\rm km~s^{-1}}$) we consider them as narrow-line Seyfert 1 \citep[NLS1; e.g.,][]{1985ApJ...297..166O,2000MNRAS.314L..17M}.
Moreover, these NLS1 show excessive X-ray emission similar to the BPT Seyfert II.
Thus, these sources can be safely discarded as SFGs.

A possible reason for this misclassification by the BPT diagram could be an overestimation of the narrow component of the Balmer lines which lowered the [\ion{N}{II}]/\ion{H}{$\alpha$} and [\ion{O}{III}]/\ion{H}{$\beta$} ratios.
However, the \citetalias{2012MNRAS.421.1043S} diagram shows its ability to correctly classify these sources accounting for their strong \ion{He}{II} emission.
Sixteen galaxies of this sample fulfill both BPT and \citetalias{2012MNRAS.421.1043S} classification schemes as SFGs simultaneously.
Additionally, we omit the highest-metallicity (12+log(O/H) $\simeq 9.2$) galaxy of these 16 SFGs which also shows a large excess of X-ray emission.
This source is at the edge of the \citetalias{2012MNRAS.421.1043S} classification curve. 
At the same time, it is located at the center of the group of this sample's NLS1s regarding metallicity (Figure \ref{fig:Lx_HeII_All}, lower left) hinting contamination by AGN activity.
This subsample of 15 sources will be considered the SFGs and the rest as AGNs for the following analysis regarding the comparisons between X-ray and \ion{He}{II} emission.

\subsection{Star-forming activity and X-ray luminosity}
\label{sec:SFR_and_X-rays}

As revealed by the SFR--$M_\star$ relation, all BPT and \citetalias{2012MNRAS.421.1043S} SFGs of this sample show intense star-forming activity with average specific SFR $\rm \left < log_{10} (sSFR/yr^{-1}) \right > \simeq -8$ with sizes from very small to average size galaxies ($ 6 \leq {\rm log_{10} (M_\star/M_\odot)} \leq 10.5$).
Most of these galaxies are dwarf and resemble Green Peas \citep[e.g.,][]{2009MNRAS.399.1191C,2011ApJ...728..161I}, or Blueberries \citep{2017ApJ...847...38Y,2024A&A...688A.159K}.
In fact, two of them are part of the known Blueberry samples \citep[PGC 140535 and 3328271;][]{2024A&A...688A.159K}. 
The AGNs, which also have higher \ion{He}{II}/\ion{H}{$\beta$}, form a separate group of higher mass host galaxies ($ 9.2 \leq {\rm log_{10} (M_\star/M_\odot)} \leq 11.8$) which are also less active with average  $\rm \left < log_{10} (sSFR/yr^{-1}) \right > \simeq -10$.
The low-SFR AGN galaxies have large uncertainties in their SFR estimations due to the overall low SFR and the contribution of the AGN in their emission which is difficult to disentangle from the star-forming activity.
Nevertheless, the SFRs of AGNs are only used in Figure \ref{fig:Lx_HeII_All} for demonstration purposes. 
They are not used in any other part of the analysis.

In the middle right plot of Figure \ref{fig:Lx_HeII_All} we compare the total $L_{\rm X}$ as a function of the SFR of the sources.
As a reference, we plot the \cite{2014MNRAS.437.1698M} relation.
Most of the \citetalias{2012MNRAS.421.1043S} SFGs follow the \cite{2014MNRAS.437.1698M} relation.
However, three sources show a significant excess of X-ray emission.
These galaxies spread over a wide range of SFRs.
All of the higher $\rm \ion{He}{II}$/\ion{H}{$\beta$} ratio, thus \citetalias{2012MNRAS.421.1043S} AGNs, are above the \cite{2014MNRAS.437.1698M} line.
Additionally, the vast majority shows significant $L_{\rm X}$ excess reaching up to four orders of magnitude. 
The average $L_{\rm X}$ excess of the \citetalias{2012MNRAS.421.1043S} AGNs is about 2~dex higher compared to that expected by their star-forming activity. 
Still, there is significant scatter.

\subsection{Normalized X-ray luminosity compared to metallicity and \texorpdfstring{\ion{He}{II}}~ emission}
\label{sec:X-rays_Z_HeIIHb}

The lower left plot of Figure \ref{fig:Lx_HeII_All} presents the normalized X-ray luminosity ($L_{\rm X}$/SFR) as a function of the metallicity (12+log(O/H)) of the sources.
This plot includes only the BPT and \citetalias{2012MNRAS.421.1043S} SFGs as MPA-JHU does not provide metallicity estimations for sources classified as AGNs by the BPT diagram.
We also plot the \cite{2014MNRAS.437.1698M} relation where the $L_{\rm X}$/SFR ratio is independent of metallicity, and the \cite{2013ApJ...776L..31F}, \cite{2016MNRAS.457.4081B} and \cite{2021ApJ...907...17L} relations which are functions of metallicity.

Most of the \citetalias{2012MNRAS.421.1043S} SFGs follow the relations mentioned above except for three low metallicity sources which show significant excess compared to what is expected by their SFR and low metallicities.
The NLS1s (BPT-SFGs-but-SB12-AGN; Section \ref{sec:classification}) form a separated group located at higher metallicities ($\rm 12+ log(O/H) > 8.8$). 
However, because of the AGN's possible contribution to the observed spectrum and the emission-line component, their metallicity estimations should be considered with caution.
All NLS1s of this sample show significantly elevated $L_{\rm X}$/SFR. 

The lower right plot of Figure \ref{fig:Lx_HeII_All} presents the $L_{\rm X}$/SFR as a function of the \ion{He}{II}/\ion{H}{$\beta$} ratio.
The \citetalias{2012MNRAS.421.1043S} SFGs and AGNs are located at the left and right parts of the diagram respectively due to the increased \ion{He}{II}/\ion{H}{$\beta$} ratio of the AGNs.
On average the AGNs also show increased $L_{\rm X}$/SFR compared to the SFGs.
The $L_{\rm X}$/SFR monotonically increases with ${\ion{\rm He}{II}}/{\rm H}\beta$ regardless of the class of the source.
A linear regression fit involving all the sources, regardless of their class, shows that there is an underlying relation that correlates the $L_{\rm X}$/SFR with the ${\ion{\rm He}{II}}/{\rm H}\beta$ ratio regardless of the nature of the X-ray emission.
However, the sources show a large scatter which is increasing at a higher ${\ion{\rm He}{II}}/{\rm H}\beta$ ratio.
The scatter is larger for the AGNs probably driven by the large uncertainties of their SFR estimations, due to their host-galaxy passive nature, and the contribution of the AGN in the galaxy's spectral energy distribution (SED) and emission-line component. 
Moreover, their X-ray emission is mainly attributed to the AGN and thus is less relevant and related to the host galaxy's SFR.

\subsection{Hardness ratio and obscuration}
\label{sec:Hardness_obscuration}

\begin{figure}[ht!]
    \centering
    \includegraphics[width=0.8\columnwidth]{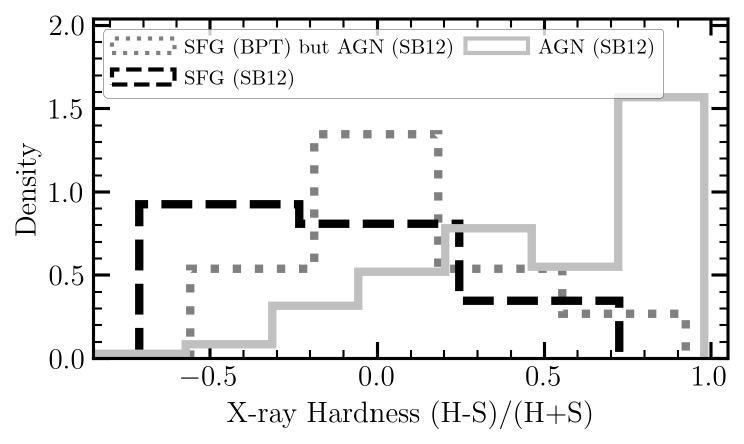}
    \includegraphics[width=0.8\columnwidth]{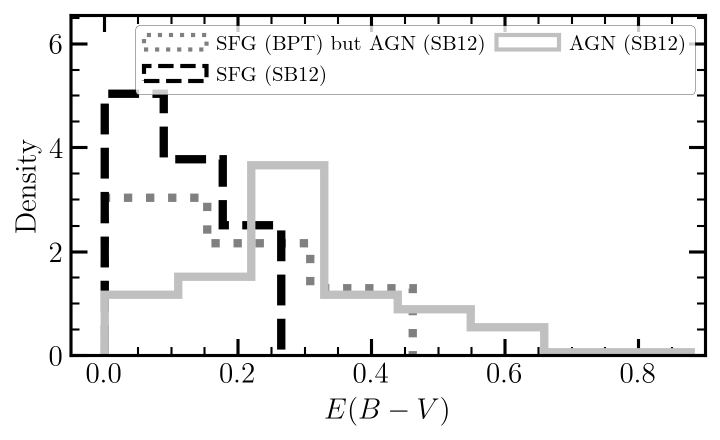}
    \caption{Histograms of the hardness ratio (top), and extinction (bottom) of our sample of \ion{He}{II} and X-ray detected sources. 
    The black dashed, and light gray continuous lines correspond to \citetalias{2012MNRAS.421.1043S} SFGs and AGNs.
    The gray dotted line represents sources classified as SFGs by BPT but AGNs by the \citetalias{2012MNRAS.421.1043S} diagram (NLS1s; Section \ref{sec:classification}).}
    \label{fig:histogramms}
\end{figure}

The distribution of the X-ray hardness for the different classes of our sample is examined in the upper plot of Figure~\ref{fig:histogramms}. 
The X-ray hardness ratio is defined here as $(H-S)/(H+S)$, where $H$ is the count rate in the hard band (2-8\,keV) and $S$ is the count rate in the soft and medium bands (0.5-2\,keV) of Chandra. 
The hardness ratio is a useful quantity that encapsulates the spectral slope of the X-ray spectrum. 
A large population of AGNs is not expected to be dominated by a very hard X-ray spectrum overall.
Obscured sources tend to have higher X-ray hardness due to absorption of the soft X-ray flux \citep[e.g.,][]{Hasinger2008, Civano2012, Fornasini2018}. 
However, the \citetalias{2012MNRAS.421.1043S} AGNs of our sample show a distribution dominated by sources with hard X-ray spectra hinting that they may be obscured sources.
On the other hand, the hardness distribution of the purely SFGs is dominated by sources with soft X-ray spectrum.
Interestingly, the hardness distribution of our sample's NLS1s stands in the middle with the distribution's mode $\rm \left < HR \right > \simeq 0$.
It should be mentioned that the homogeneous analysis by CSC does not specifically fit the X-ray spectra of the sources with obscured models so the delivered fluxes do not account for intrinsic obscuration.

The bottom panel of Figure~\ref{fig:histogramms} presents the extinction of the sources measured in $E(B-V)$.
The $E(B-V)$ was estimated spectroscopically through the Balmer ratio assuming case-B recombination and the reddening law of \cite{2000ApJ...533..682C}.
Similarly to the hardness ratio, the extinction is significantly higher for the Seyfert II (BPT and \citetalias{2012MNRAS.421.1043S}) AGNs compared to the SFGs and NLS1s.
The modes of their $E(B-V)$ distribution lie at 0.05, 0.05, and 0.27 for the SFGs, NLS1s, and Seyfert II AGNs respectively, showcasing the the latter's dustier nature.
Our sample's Seyfert II AGNs show hard spectra and high reddening, a combination suggesting that they are obscured sources. 

\subsection{The relation of normalized X-ray luminosity and \texorpdfstring{\ion{He}{II}}{} to \texorpdfstring{\ion{H}{$\beta$}}{} ratio of purely star-forming galaxies}
\label{sec:SFGs_LXSFR_HeIIHb}

\begin{figure*}
    \centering
    \includegraphics[width=0.98\textwidth]{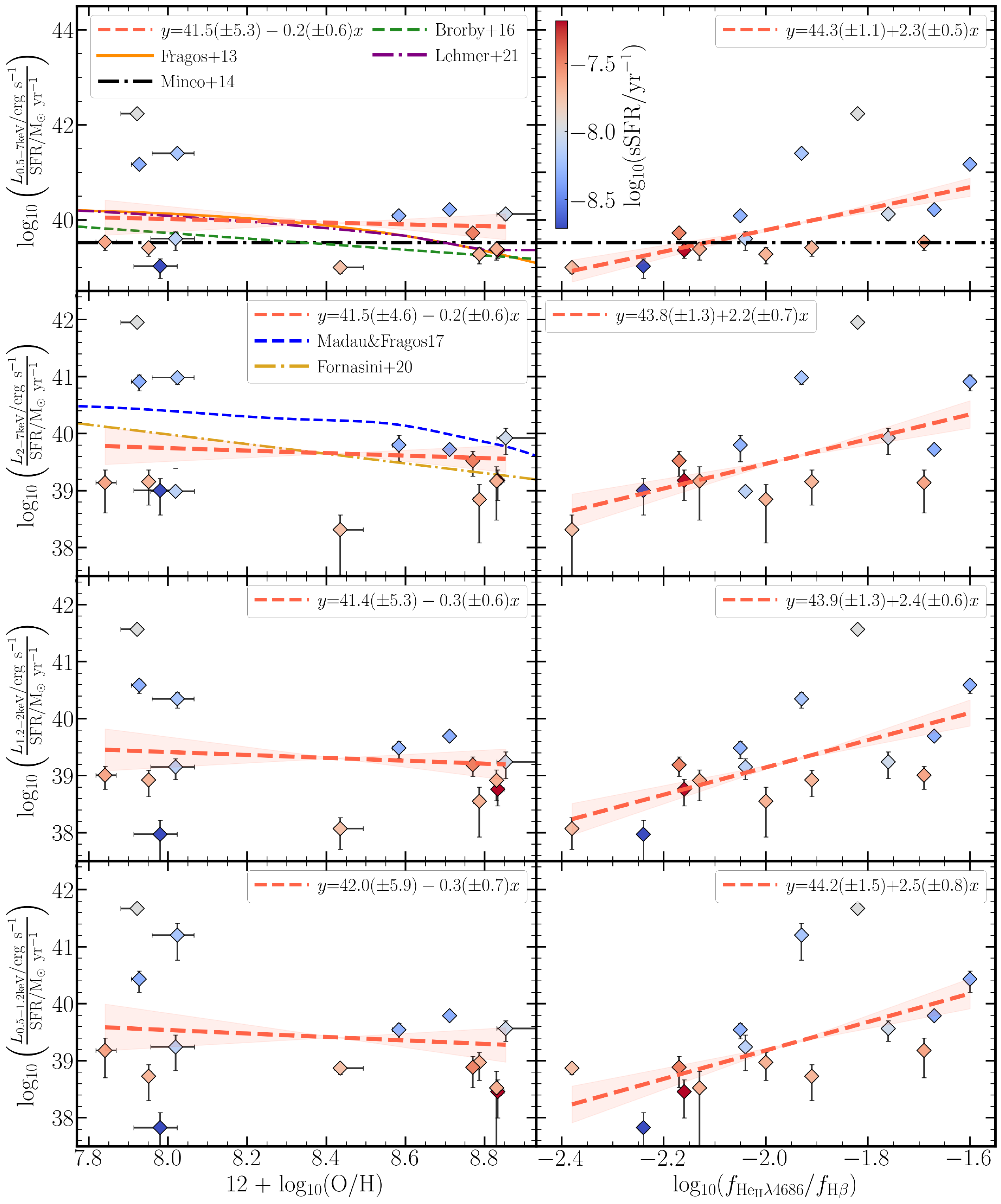}
    \caption{$L_{\rm X}$/SFR as a function of metallicity (left column), or \ion{He}{II}/\ion{H}{$\beta$} ratio (right column) of SFGs classified by both the \citetalias{2012MNRAS.421.1043S} and BPT diagrams.
    The energy bands total, hard, medium, and soft appear from top to bottom respectively. 
    All points are color-coded based on the galaxies' sSFR.
    The red dashed line represents the best linear regression fit and the shaded area shows the uncertainties of the fit.
    The orange continuous, black dashed-doted line, green dashed, and purple dashed-dotted curves represent the \cite{2013ApJ...776L..31F}, \cite{2014MNRAS.437.1698M}, \cite{2016MNRAS.457.4081B}, and \cite{2021ApJ...907...17L} relations respectively.
    The blue dashed and the yellow dashed-dotted lines shown in the second row correspond to the \cite{2017ApJ...840...39M} and \cite{2020MNRAS.495..771F} relations respectively.}
    \label{fig:Lx_HeII_SFGs}
\end{figure*}

In this section, we examine more closely the $L_{\rm X}$/SFR--{\ion{\rm He}{II}}/\ion{H}{$\beta$} relation for our sample's SFGs.
In Figure \ref{fig:Lx_HeII_SFGs} we present the relation between the $L_{\rm X}$/SFR and metallicity (left column), or ${\ion{\rm He}{II}}/{\rm H}\beta$ ratio (right column) in logarithmic space for the 15 purely SFGs of our sample based on the \citetalias{2012MNRAS.421.1043S} and BPT classification.
The comparison is performed separately for the CSC's total, hard, medium, and soft X-ray energy bands while all points are color-coded based on the galaxies' sSFR.

The galaxies form two groups of almost the same number of sources regarding metallicity. 
At the left are low metallicity sources with 12+log(O/H) $\lessapprox 8$, and at the right are sources scattered around the solar metallicity ($\rm 8.4 < 12+log(O/H) < 8.9$).
We also compare with the established \cite{2014MNRAS.437.1698M}, and the \cite{2013ApJ...776L..31F}, \cite{2016MNRAS.457.4081B}, and \cite{2021ApJ...907...17L} relations where the $L_{\rm X}$/SFR is a function of metallicity.
The low-metallicity sources show a large scatter above and below these relations.
The observed scatter can be attributed to the stochastic nature of XRBs or the presence of ULXs.
Both scenarios can have a stronger effect on low-mass galaxies.
Conversely, the near-solar metallicity sources tend to follow the reference relations more closely.

The comparison between the $L_{\rm X}$/SFR and ${\ion{\rm He}{II}}/{\rm H}\beta$ ratios (right column) shows a positive correlation which holds in all the examined energy bands.
For a given $L_{\rm X}$/SFR, the observed \ion{He}{II}/\ion{H}{$\beta$} ratios of these SFGs align exclusively with the highest X-ray efficiency models proposed by \cite{2020MNRAS.494..941S}. 
Notably, these models predict a strong positive correlation between $L_{\rm X}$/SFR and ${\ion{\rm He}{II}}/{\rm H}\beta$, which is consistent with the findings presented in this study.
Galaxies with low ${\ion{\rm He}{II}}/{\rm H}\beta$ show low X-ray emission and vice versa, sources with high ${\ion{\rm He}{II}}/{\rm H}\beta$ show higher $L_{\rm X}$/SFR.
We perform a linear regression fit to quantify this correlation. 
The slope of the linear regression fits is higher for the soft CSC band but the results are similar for all bands accounting for the fit uncertainties.
This agrees with the fact that most of the SFGs of our sample show a rather soft X-ray spectrum (Section \ref{sec:Hardness_obscuration}).
Overall, the ${\ion{\rm He}{II}}/{\rm H}\beta$ ratio appears as an enhanced tracer of the $L_{\rm X}$/SFR ratio compared to metallicity showing lower scatter and steeper slopes.

The majority of the SFGs closely follow the regression fits, besides two sources with significant X-ray excess.
The color-coding of Figure \ref{fig:Lx_HeII_SFGs} reveals that most intensively star-forming sources $\rm log~(sSFR/yr^{-1}) \leq -8$ tend to be below these linear regression fits.
On the contrary, the ones slightly less actively star-forming, with $\rm log~(sSFR/yr^{-1}) \geq -8$, tend to show an excess of $L_{\rm X}$/SFR.
As the sSFR is the ratio of the SFR over the $M_\star$ of the galaxies, it traces the youth and starburst intensity of their SPs.
All these sources host very young SPs with $\rm log~(sSFR/yr^{-1}) \leq -9$.
However, the fact that those with the highest sSFR show relatively lower $L_{\rm X}$/SFR indicates a dependence on the star-forming activity and possibly the age of the SPs. 

Observations of the nearby Large and Small Magellanic Clouds have shown that some exceptionally young SPs show a deficit in X-ray emission  \citep{2010ApJ...716L.140A,2019ApJ...887...20A}.
This has also been discussed as the reason for the X-ray luminosity deficit of some of the highly star-forming dwarf Blueberry galaxies \citep[e.g.,][]{2024A&A...691A..27A}.
Models of HMXBs showed that very young SPs, with ages less than 5 Myr host fewer HMXBs and thus, show lower X-ray intensity \citep[e.g.,][]{2007AstL...33..437S} while for SPs with ages below 3 Myrs, they may be completely absent \citep{2024ApJ...977..189L}.
This deficit is attributed to the ages of their SPs and the time needed to produce XRBs.
Very young galaxies may show lower X-ray emission compared to those with slightly older SPs since the peak of their X-ray emission is estimated to be around 30--50~Myrs \citep[e.g.,][]{2013ApJ...776L..31F}. 
Our sample tends to follow this trend.

The total X-ray output of a galaxy includes the XRBs and the circumgalactic medium's hot gas emission.
While the hot gas' contribution can be significant in passive galaxies, it becomes less important in starburst galaxies like this samples' SFGs \citep[e.g.,][]{2022ApJ...926...28G,2023A&A...669A..34V,2025A&A...695A...2V}.
Recent analysis on a large sample of X-ray observed sources estimated that, on average, the contribution of hot gas in the total X-ray output of SFGs is around 10\% \citep{2025A&A...694A.128K}.
Moreover, the majority of our sample's SFGs are dwarf galaxies with $M_\star < 10^{9.5} M_\odot$ which are known not to be able to sustain large amounts of hot gas due to their weaker gravitational fields \citep[e.g.,][]{2011ApJ...729...12B,2012MNRAS.426.1870M}.
Thus, we do not expect strong biases in the adopted fluxes due to hot-gas contribution.

\subsection{Direct comparison between X-ray and \texorpdfstring{\ion{He}{II}}{} luminosities}
\label{sec:X_ray_HeII_luminosities}

Since the SFR is correlated and can be estimated through the strength of the Balmer emission lines \citep[e.g.,][]{1998ARA&A..36..189K}, we can remove the denominators of both terms of the $L_{\rm X}$/SFR--$\rm \ion{He}{II}$/\ion{H}{$\beta$} relation and examine directly the relation between the X-ray and $\rm \ion{He}{II}$ luminosities.
The work of \citetalias{2012MNRAS.421.1043S} does not provide the \ion{He}{II} luminosities.
We recover them through the given $\rm \ion{He}{II}$/\ion{H}{$\beta$} ratios and the \ion{H}{$\beta$} emission-line fluxes estimated by MPA-JHU which are based on the same SDSS spectra.

Moreover, since the X-ray emission corresponds to the complete spatial extent of the sources, we need to correct the cases where the SDSS fiber does not fully cover the sources.
To perform aperture correction on sources that may be larger than the SDSS fiber (3\arcsec~box), we crossmatch our sample with the Heraklion Extragalactic Catalog \citep[HECATE;][]{2021MNRAS.506.1896K}\footnote{\url{https://hecate.ia.forth.gr/}} that provides the angular sizes of galaxies up to 200 Mpc.
We find that only three sources of our sample have a major axis larger than 3\arcsec, for which we multiply the ratio of their major axis over the size of the SDSS fiber with the \ion{He}{II} luminosity derived by the SDSS spectra.
The aperture correction may suffer from biases due to the potential spatial variation of the \ion{He}{II} emission on the surface of the galaxy.
However, since the correction is required only for 3/165 sources we do not expect it to significantly affect the results of this analysis. 

\begin{figure*}[ht!]
    \centering
    \includegraphics[width=1\textwidth]{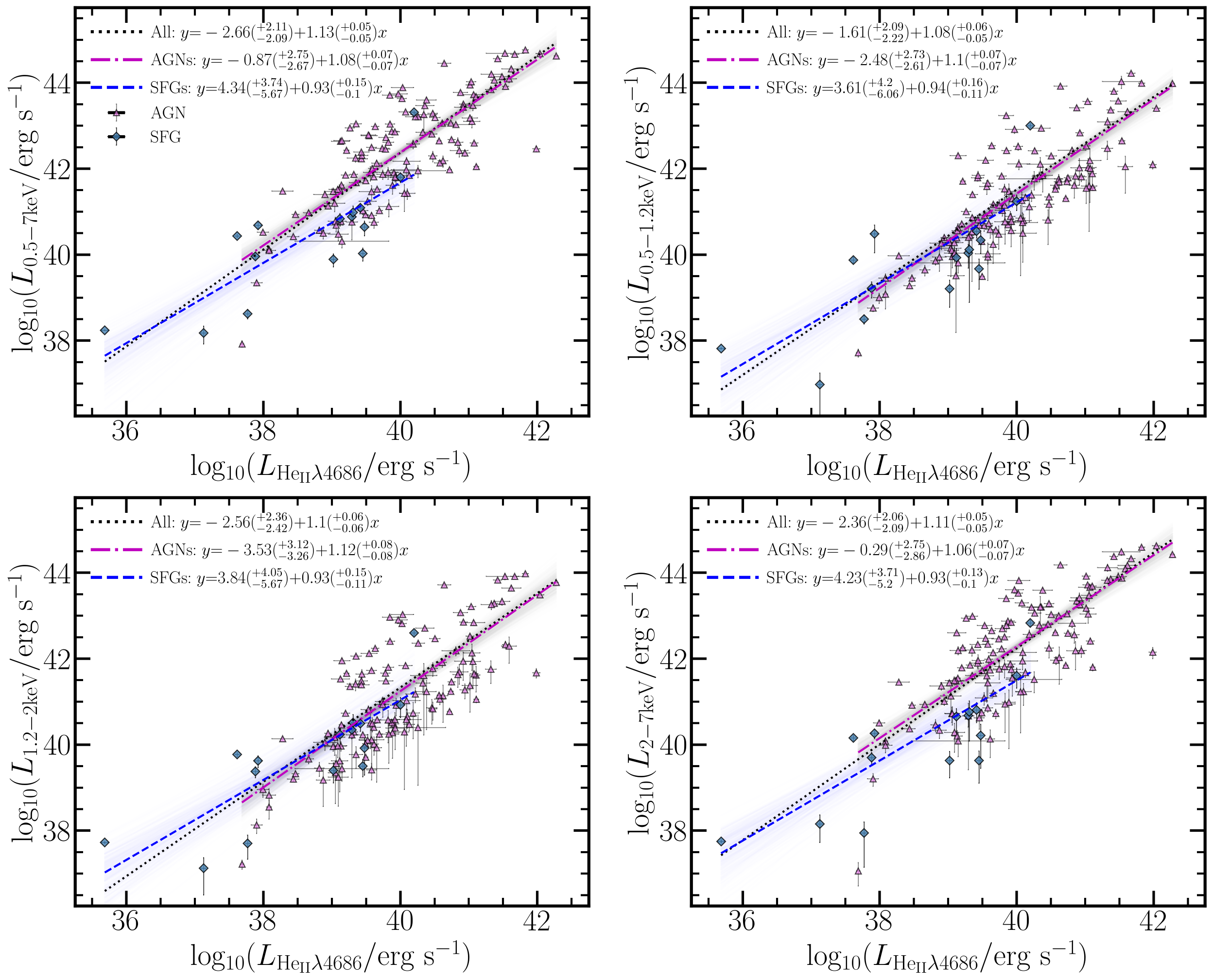}
    \caption{X-ray as a function of $\ion{\rm He}{II}$ luminosity for different X-ray energy bands in logarithmic space.
    The energy bands total, soft, medium, and hard appear from top to bottom and left to right respectively. 
    The pink triangles and blue rhombuses show AGNs and SFGs respectively as adopted by this work (Section \ref{sec:classification}).
    The black dotted, purple dashed-dotted, and blue dashed lines correspond to the best results of MCMC fits involving all sources, AGNs, and SFGs respectively.
    The shaded areas with the same colors correspond to the uncertainties of the respective fits.
    The results of the MCMC fits are presented at the top of each plot and are summarized in Table \ref{tab:Lx_LHeII_linear_fit}.}
    \label{fig:Lx_HeII_bands}
\end{figure*}

The \ion{He}{II}--X-ray luminosity relation is examined in Figure \ref{fig:Lx_HeII_bands} separately for the provided soft, medium, hard, and total CSC energy bands.
To quantify the correlation we performed Markov-chain Monte-Carlo (MCMC) fits adopting a linear relation between the X-ray and \ion{He}{II} luminosities. 
The relation is in the form:  $\rm log_{10} \left ( \frac{\textit{L}_{X}}{erg~s^{-1}} \right ) = \alpha + \beta ~ log_{10} \left ( \frac{\textit{L}_{\ion{He}{II}}}{erg~s^{-1}} \right ) + \sigma $ where the $\alpha$, $\beta$, and $\sigma$ represent the intercept, the slope, and the variance respectively.
The MCMC fitting was performed with the \texttt{Python emcee} package \citep[][]{emcee}.
The fitting implemented 32 walkers over 5000 iterations with the first 500 for the burn-in phase.
To estimate the initialization parameters, we performed a maximum likelihood fit with the minimization algorithm \texttt{scipy} \citep{2020SciPy-NMeth}.
We adopted a uniform prior throughout the process.
The results of the MCMC fitting are provided separately for the different X-ray energy bands in Table \ref{tab:Lx_LHeII_linear_fit}.
As represented by the shaded areas in Figure \ref{fig:Lx_HeII_bands}, the uncertainties of the fit are quite low accounting for the fact that the intercepts and slopes of each fit are strongly anti-correlated.

\begin{table*}[ht!]
    \renewcommand{\arraystretch}{1.5}
    \centering
    \caption{Summary of MCMC fitting results of the X-ray and \ion{He}{II} correlation.}
    \begin{threeparttable}
    \begin{tabular}{cccccc}
        Energy (keV) & Class & $\alpha$ & $\beta$ & $\rm log_{10} (\sigma$) & $\rm std~ \left[log_{10} \left( \frac{\textit{L}_{X}}{\textit{L}_{\ion{He}{II}}} \right) \right]$ \\
        \hline
        0.5--7 & All & $-2.66^{+2.11}_{-2.09}$ & $1.13^{+0.05}_{-0.05}$ & $-4.09^{+0.06}_{-0.05}$ & 0.72\\
        0.5--7 & AGNs & $-0.87^{+2.75}_{-2.67}$ & $1.08^{+0.07}_{-0.07}$ & $-4.08^{+0.06}_{-0.06}$ & 0.72\\
        0.5--7 & SFGs & $4.34^{+3.74}_{-5.67}$ & $0.93^{+0.15}_{-0.1}$ & $-3.92^{+0.22}_{-0.19}$ & 0.75\\
        \hline
        0.5--1.2 & All & $-1.61^{+2.09}_{-2.22}$ & $1.08^{+0.06}_{-0.05}$ & $-4.06^{+0.06}_{-0.06}$ & 0.75\\
        0.5--1.2 & AGNs & $-2.48^{+2.73}_{-2.61}$ & $1.1^{+0.07}_{-0.07}$ & $-4.1^{+0.07}_{-0.07}$ & 0.7\\
        0.5--1.2 & SFGs & $3.61^{+4.2}_{-6.06}$ & $0.94^{+0.16}_{-0.11}$ & $-3.86^{+0.25}_{-0.22}$ & 0.85\\
        \hline
        1.2--2 & All & $-2.56^{+2.36}_{-2.42}$ & $1.1^{+0.06}_{-0.06}$ & $-3.95^{+0.06}_{-0.06}$ & 0.82\\
        1.2--2 & AGNs & $-3.53^{+3.12}_{-3.26}$ & $1.12^{+0.08}_{-0.08}$ & $-3.92^{+0.07}_{-0.07}$ & 0.82\\
        1.2--2 & SFGs & $3.84^{+4.05}_{-5.67}$ & $0.93^{+0.15}_{-0.11}$ & $-3.94^{+0.23}_{-0.22}$ & 0.76\\
        \hline
        2--7 & All & $-2.36^{+2.06}_{-2.09}$ & $1.11^{+0.05}_{-0.05}$ & $-4.11^{+0.06}_{-0.06}$ & 0.77\\
        2--7 & AGNs & $-0.29^{+2.75}_{-2.86}$ & $1.06^{+0.07}_{-0.07}$ & $-4.06^{+0.07}_{-0.07}$ & 0.77\\
        2--7 & SFGs & $4.23^{+3.71}_{-5.2}$ & $0.93^{+0.13}_{-0.1}$ & $-4.08^{+0.26}_{-0.25}$ & 0.75\\
    \end{tabular}
    \begin{tablenotes}
    \item[note 1]The adopted MCMC model is: $\rm log_{10} \left ( \frac{\textit{L}_{X}}{erg~s^{-1}} \right ) = \alpha + \beta ~ log_{10} \left ( \frac{\textit{L}_{\ion{He}{II}}}{erg~s^{-1}} \right ) + \sigma $.
    \end{tablenotes}
    \end{threeparttable}
    \label{tab:Lx_LHeII_linear_fit}
\end{table*}

The comparisons and fits presented in Figure \ref{fig:Lx_HeII_bands} display a tight linear correlation between X-ray and \ion{He}{II} luminosities regardless of the class of the sources.
This correlation holds for all X-ray energy bands, including the total, soft, medium, and hard, extending for about seven orders of magnitude.
On average, the X-ray luminosity is about 2.5 orders of magnitude higher than that of the \ion{He}{II}.
Most of the low $L_{\rm \ion{He}{II}}$ and $L_{\rm X}$ sources are purely SFGs while the most luminous are AGNs.
In addition to the hard X-ray band where the fit corresponding to the SFGs is slightly lower than the respective fit of the AGNs, their fits overlap, indicating a similar excitation mechanism for both classes.
The displayed scatter between the luminosities is overall lower for the total X-ray emission.
The SFGs show a larger scatter compared to AGNs, a fact which however could be attributed to the relatively smaller size of the sample.

\section{Investigation of the ionization properties of all (agnostic to X-rays) \texorpdfstring{\ion{He}{II}}{} galaxies}
\label{sec:All_HeII}

\begin{figure*}[ht!]
    \centering
    \includegraphics[width=0.98\textwidth]{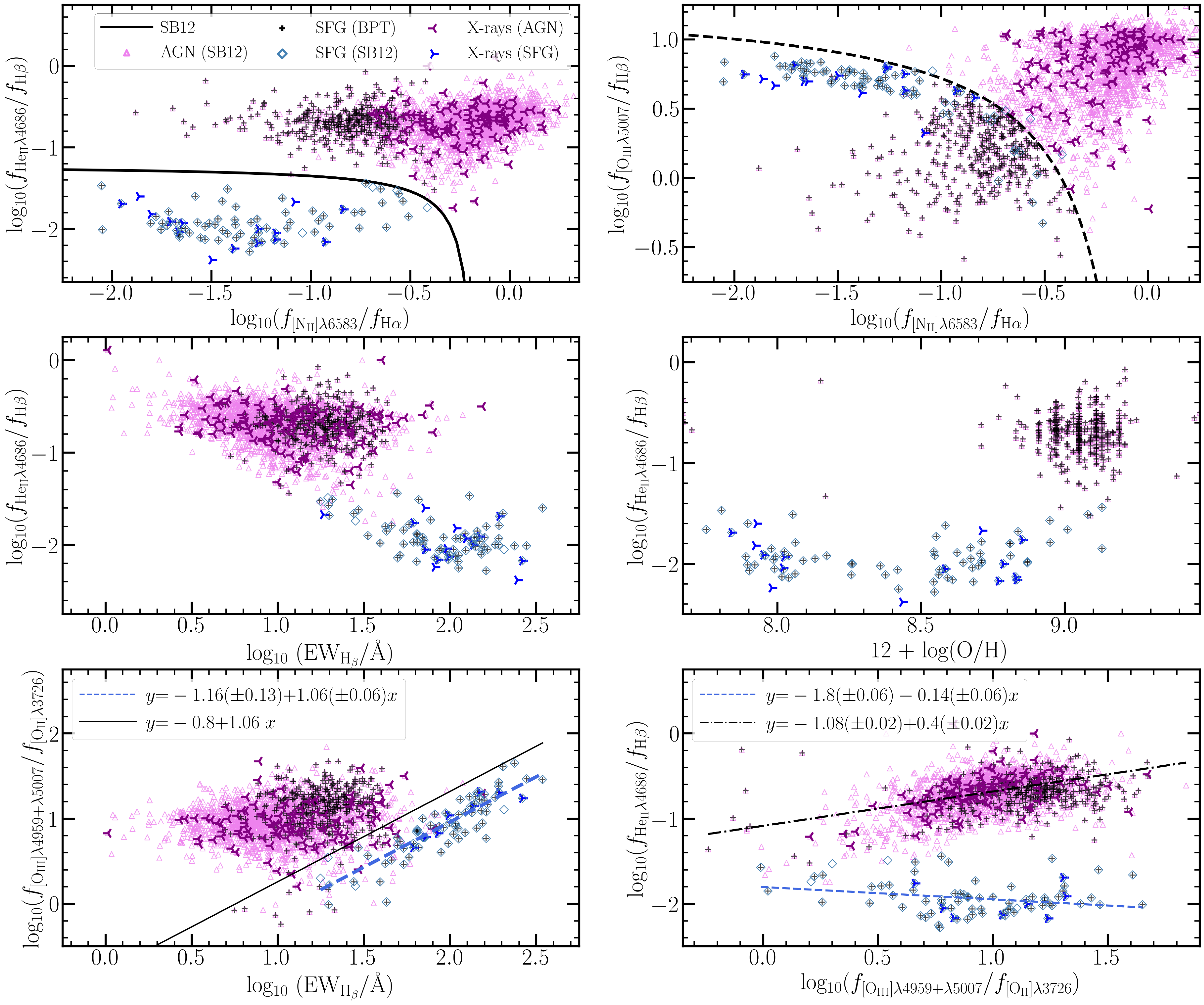}
    \caption{Several comparisons of 2399 galaxies emitting $\rm \ion{He}{II}$ with  $\rm SnR>5.5$, measured and cataloged by \citetalias{2012MNRAS.421.1043S}, and matched with the SDSS MPA-JHU catalog having $\rm S/N > 5$ in all compared emission lines. 
    \citetalias{2012MNRAS.421.1043S} SFGs and AGNs are represented by light blue rhombuses and violet triangles respectively.
    BPT SFGs are represented by a black cross.
    The X-ray-selected SFGs and AGNs (Section \ref{sec:classification}) are also shown with blue right-pointing and purple left-pointing triangles respectively.
    Top left: \ion{He}{II}/\ion{H}{$\beta$} over [\ion{N}{II}]/\ion{H}{$\alpha$}. 
    The black curve represents the \citetalias{2012MNRAS.421.1043S} classification criterion separating AGN from SFGs.
    Top right: The BPT [\ion{O}{III}]/\ion{H}{$\beta$} over [\ion{N}{II}]/\ion{H}{$\alpha$} diagram. 
    The dashed black line shows the 
    \cite{2003MNRAS.346.1055K} classification criterion.
    Middle left: \ion{He}{II}/\ion{H}{$\beta$} as a function of $\rm EW_{H\beta}$.
    Middle right:  \ion{He}{II}/\ion{H}{$\beta$} as a function of metallicity.
    Bottom left: O32 as a function of $\rm EW_{H\beta}$.
    The blue dashed line represents the best linear regression fit involving only BPT and \citetalias{2012MNRAS.421.1043S} SFGs.
    The black line represents a threshold above which only AGNs of our sample are located.
    Bottom right: \ion{He}{II}/\ion{H}{$\beta$} as a function of O32.
    The black dashed-dotted and blue dashed lines show the best linear regression fits involving solely \citetalias{2012MNRAS.421.1043S} AGNs or SFGs, respectively. 
    }
    \label{fig:HeII_OIII_OII}
\end{figure*}

This section expands on the former analysis and investigates some of the ionization properties of all the \citetalias{2012MNRAS.421.1043S} $\ion{He}{II}$ galaxies regardless of whether they have been detected in X-rays or not.
In Figure \ref{fig:HeII_OIII_OII}, we compare several parameters of the \citetalias{2012MNRAS.421.1043S} sample.
The comparisons galaxies emitting $\rm \ion{He}{II}$ with  $\rm SnR>5.5$, measured and cataloged by \citetalias{2012MNRAS.421.1043S}, and matched with the SDSS MPA-JHU catalog.
We omit sources having $\rm S/N < 5$ in the $\rm [\ion{O}{II}]\lambda 3726$, $\rm [\ion{O}{III}]\lambda 4959$, $\rm [\ion{O}{III}]\lambda 5007$, and \ion{H}{$\beta$} emission lines yielding a sample of 2399 galaxies.
We also plot the X-ray-selected sample (Section \ref{sec:classification}).
The larger, agnostic-to-X-rays sample partially serves as a control sample for the sources with detected X-ray and \ion{He}{II} emission.
We note that some X-ray-selected sources lack $\rm [\ion{O}{II}]\lambda 3726$ detection because this line can fall outside the SDSS spectrum range due to their higher redshift.
Conversely, some X-ray sources do not have counterparts in the agnostic sample because of the S/N constraints applied to all the emission lines used in these comparisons.
The X-ray-selected AGNs and SFGs exhibit distributions similar to their counterparts in the agnostic sample. 
This suggests that the X-ray-detected sample is representative and not significantly different from the majority of \ion{He}{II} emitters.

\begin{figure}[ht!]
    \centering
    \includegraphics[width=0.8\columnwidth]{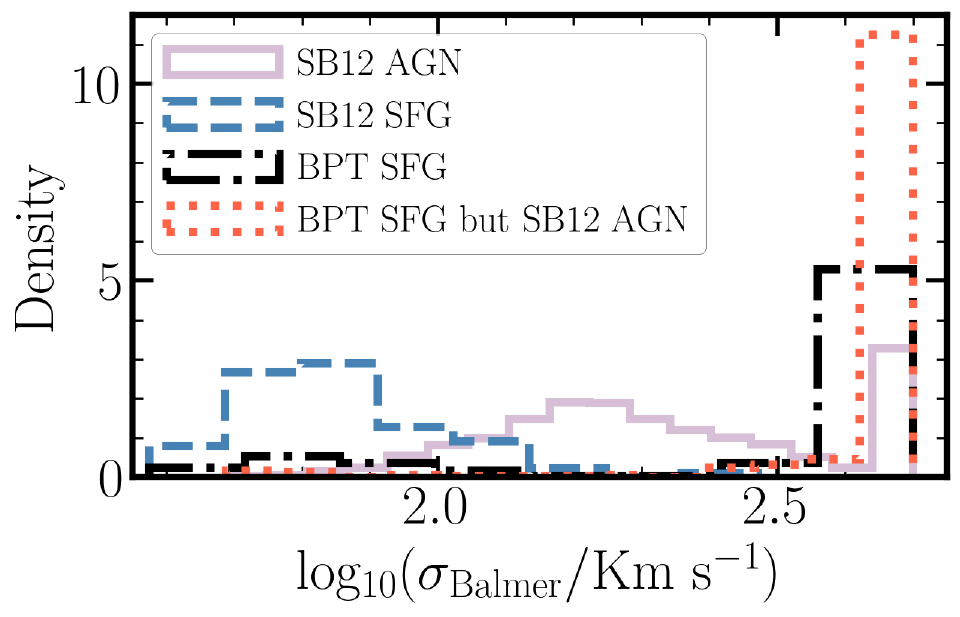}
    \caption{Histograms of the logarithm of the Balmer lines velocity dispersion ($\sigma$) for the different classes of the agnostic-to-X-rays sample.
    The pink continuous and blue dashed lines show the AGNs and SFGs classified by the \citetalias{2012MNRAS.421.1043S} diagram.
    The SFGs based on the BPT diagram are presented with a black dashed-dotted line while the red dotted line shows those classified as SFGs by the BPT but as AGNs by the \citetalias{2012MNRAS.421.1043S} diagram.}
    \label{fig:HeII_O32_Sigma_Hist}
\end{figure}

The first two plots show the BPT and the \citetalias{2012MNRAS.421.1043S} diagrams, providing classification for the sources.
The BPT diagram uses optical emission-line ratios to distinguish between AGNs and SFGs, while the \citetalias{2012MNRAS.421.1043S} diagram provides an alternative classification based on \ion{He}{II}/\ion{H}{$\beta$} flux ratios. Comparing the two helps identify discrepancies in classification methods and provides insight into the ionization mechanisms at play.
Similarly to the X-ray sample, a significant number of sources are considered SFGs by the BPT but as AGNs by \citetalias{2012MNRAS.421.1043S}.
Based on \citetalias{2012MNRAS.421.1043S}, 2322 galaxies are classified as AGNs and 77 as SFGs.
Based on the BPT 431 galaxies are located below the \cite{2003MNRAS.346.1055K} line and are classified as purely SFGs.
As discussed in Section \ref{sec:classification} most of the BPT SFG but \citetalias{2012MNRAS.421.1043S} AGNs of our X-ray sample show broadened emission lines indicating contribution from the BLR of an AGN.
Histograms of the Balmer-lines velocity dispersion ($\rm \sigma_{Balmer}$) of this larger sample are presented in Figure \ref{fig:HeII_O32_Sigma_Hist} separated to the SFG and AGN classes based on the different methods.
Once again it is shown that the sources identified as SFGs by the BPT but as AGNs by \citetalias{2012MNRAS.421.1043S} are those showing the largest $\sigma$ among the galaxies of this sample indicating the presence of an AGN.
Therefore, for the following comparisons and discussion, we consider purely star-forming sources classified as such by both diagrams.
The rest are considered as AGNs.

The middle left plot of Figure \ref{fig:HeII_OIII_OII} shows the \ion{He}{II}/\ion{H}{$\beta$} ratio as a function of the equivalent width of the \ion{H}{$\beta$} line ($\rm EW_{H\beta}$) in logarithmic space.
The $\rm EW$ corresponds to the ratio of the emission or absorption component of the line over the continuum.
Because the continuum is mainly attributed to the thermal emission of the long-lived low-mass stars it traces the strength and presence of the old SPs. 
The spectral line component, when in emission, traces the gas excitation from the young SPs.
Thus, the EW is an excellent tracer of the SPs' youth and the galaxies' star-forming activity.
Therefore, the EWs of the Balmer lines show a strong correlation with the sSFR \citep[e.g.,][]{2018MNRAS.477.3014B}.

We observe the two classes aggregating in two clusters, with the AGNs at the top left and the SFGs at the bottom right.
The SFGs have high $\rm EWs_{H\beta}$ but show low \ion{He}{II}/\ion{H}{$\beta$} ratios.
This lack of correlation indicates that the radiation of the young SPs, which drive the $\rm EWs_{H\beta}$, can not be the excitation mechanism behind \ion{He}{II} emission.
Conversely, the host galaxies of most AGNs are passive with no significant star-forming activity as shown by their low $\rm EWs_{H\beta}$. 
However, they show a large scatter in the comparison with the $\rm EWs_{H\beta}$, and higher \ion{He}{II}/\ion{H}{$\beta$} ratios which can be attributed to the AGN's activity.

The middle right plot of Figure \ref{fig:HeII_OIII_OII} shows the \ion{He}{II}/\ion{H}{$\beta$} ratio as a function of metallicity including all BPT SFGs for which MPA-JHU provides metallicity estimations. 
The majority of the BPT-SFGs-but-SB12-AGNs form a cluster of points with high metallicities and high \ion{He}{II}/\ion{H}{$\beta$} ratios. 
As discussed above these sources are probably AGNs, thus, their metallicity estimations should be considered with caution. 
The pure SFGs of our sample cover a large range of metalicities within $\rm 7.7 \lesssim 12 + log(O/H) \lesssim  9.2 $.
There is no strong correlation found between the \ion{He}{II}/\ion{H}{$\beta$} ratio and the metallicity of the SFGs across the full metallicity range.
However, the sources tend to show a bimodal behavior albeit their substantial scatter.
Galaxies with $\rm 12 + log(O/H) >  8.5 $ show a slightly positive correlation with increased \ion{He}{II}/\ion{H}{$\beta$} at super-solar metallicities while low metallicity SFGs with $\rm 12 + log(O/H) < 8.5 $ tend to show an anti-correlation. 

The bottom left plot of Figure \ref{fig:HeII_OIII_OII} presents the comparison between the ${\rm log_{10}}(f_{[\ion{\rm O}{III}] \lambda \lambda 4959,5007}/f_{[\ion{\rm O}{II}] \lambda 3726})$ ratio (O32) and $\rm EW_{H\beta}$.
The O32 primarily measures the ionization state of the gas and is mainly used to estimate a) the ionization parameter $U$, which describes the number of ionizing photons per hydrogen atom in the ionized gas, and b) the effective temperature ($T_{\rm eff}$) of the ionizing stars, particularly the hot, young stars in \ion{H}{II} regions \citep[e.g.,][]{2016ApJ...816...23S}.
Moreover, it has been used as a method to distinguish between galaxies dominated by AGN activity and those dominated by star formation \citep[e.g.,][]{2002ApJS..142...35K} and it has been proposed as a tool to estimate the escape fraction of ionizing photons \cite[e.g.,][]{2014MNRAS.442..900N}.

The O32 to $\rm EW_{H\beta}$ comparison also shows a separate behavior between the two classes.
The AGNs are distributed with rather large scatter at high O32 values with most showing low $\rm EWs_{H\beta}$, confirming that the stellar activity can not be the excitation mechanism in AGNs.
However, the two parameters show a very tight linear correlation for purely SFGs.
\cite{2015A&A...576A..83S} has also reported this behavior on purely star-forming and high $U$ galaxies.
They suggested that the observed correlation is due to the increase of the effective temperature ($T_{\rm eff}$) of the interstellar medium (ISM) by ionizing (young) SP's radiation, while it is also against Lyman continuum escape which would otherwise lead to a drop of the $\rm EW_{H\beta}$.

We perform a linear regression fit involving only the SFGs of this sample to quantify the observed correlation.
The fit results to ${\rm log_{10}}(f_{[\ion{\rm O}{III}] \lambda \lambda 4959,5007}/f_{[\ion{\rm O}{II}] \lambda 3726}) = -1.16 (\pm 0.13) + 1.06 (\pm 0.06) ~ \rm log_{10}~(EW_{H\beta}/$\AA).
The standard deviation (std) of the sources around the best-fit line is only 0.18 dex confirming the very tight correlation. 
Based on the best fit and the scatter of the SFGs around it we draw a line ${\rm log_{10}}(f_{[\ion{\rm O}{III}] \lambda \lambda 4959,5007}/f_{[\ion{\rm O}{II}] \lambda 3726}) = -0.8 + 1.06 ~ \rm log_{10}~(EW_{H\beta}/$\AA) above which no SFGs of our sample are located.
However, it should be noted that some AGNs reside under this line.

The bottom right plot of Figure \ref{fig:HeII_OIII_OII} presents the \ion{He}{II}/\ion{H}{$\beta$} ratio as a function of O32.
The \ion{He}{II}/\ion{H}{$\beta$} ratio increases monotonically with O32 only for the AGNs.
A linear regression fit of the AGN sources captures this positive correlation.
However, it is probably a simplified approach since we observe a flattening at very high O32 values while the fit's slope is not so steep.
This correlation indicates that the dominating ionization source in AGN, the central super-massive BH, also affects the ionization state of the host-galaxy gas.
On the other hand, there is no obvious correlation between the \ion{He}{II}/\ion{H}{$\beta$} and O32 ratios in SFGs.
The respective linear regression fit shows a slightly negative slope, but it could be considered flat, accounting for the fit's uncertainties and the source's scatter.
This comparison reveals that the \ion{He}{II}/\ion{H}{$\beta$} and O32 ratios are uncorrelated across SFGs emitting \ion{He}{II}.

\section{Discussion}
\label{sec:discussion}

\subsection{Wolf-Rayet stars as source of \texorpdfstring{\ion{He}{II}}{} excitation}

WR stars show \ion{He}{II} emission in their spectra and thus are prime candidates for producing this emission in galactic scales. 
WR stars tend to be more abundant in galaxies with relatively higher metallicity due to various factors related to stellar evolution. 
These evolved massive stars are susceptible to metallicity.
Mainly due to decreased mass loss through stellar winds in low-metallicity environments, the minimum mass for forming WR stars is lower at higher metallicities while their lifetimes become longer
\citep[e.g.,][]{1980A&A....90L..17M,1991A&A...242...93M, 1991A&A...244..273M}. 
It has also been shown that short bursts of star formation favor the formation of WR stars \citep[e.g.,][]{1995A&A...298..767M}.
Moreover, WR stars are typically linked with young, massive star formation, which tends to occur in galaxies with higher SFR. 
Low-SFR or passive galaxies, generally show fewer WR stars because these stars are short-lived and emerge in environments with significant ongoing star formation \citep[e.g.,][]{1991A&A...244..273M}.

\cite{2013ApJ...766...91J} modeled the photoionization around hot WR stars using CLOUDY \citep[][]{Ferland_1998}. 
Their modeling showed that only the hottest WR stars could show ${\rm log_{10}}(f_{\ion{\rm He}{II}}/f_{{\rm H}\beta}) \simeq -1$.
However, conditions leading to such \ion{He}{II}/\ion{H}{$\beta$} ratios would also be combined with ${\rm log_{10}}(f_{[\ion{\rm O}{III}] \lambda \lambda 4959,5007}/f_{[\ion{\rm O}{II}] \lambda 3726}) > 1$.
Their work concludes that only 4--5~Myr starbursts, containing a substantial WR population, can simultaneously match the O32 and \ion{He}{II}/\ion{H}{$\beta$} ratios.
In this work, none of our samples' SFGs reach ${\rm log_{10}}(f_{\ion{\rm He}{II}}/f_{{\rm H}\beta}) > -1.2$ while some sources show ${\rm log_{10}}(f_{[\ion{\rm O}{III}] \lambda \lambda 4959,5007}/f_{[\ion{\rm O}{II}] \lambda 3726}) \simeq 1.7$.
Moreover, as shown in Figure \ref{fig:HeII_OIII_OII}, there is a lack of correlation between these parameters. 
Thus, they do not match these extreme conditions which would allow WR stars to explain the observed ratios.

Similarly, the analysis of  \citetalias{2012MNRAS.421.1043S} showed that only 4--5~Myr starbursts with large WR populations can reproduce the hard ionizing UV continuum required for the observed \ion{He}{II}/\ion{H}{$\beta$} ratios.
However, they discuss that binary stellar evolution allows for a larger fraction of WR stars at a wider age range.
Moreover, their dedicated spectral analysis showed that a large fraction of their sample does not show WR features, a fraction that increases with decreasing metallicity to reach about 70\% at 12+log(O/H)~$<8.2$. 
Therefore, \citetalias{2012MNRAS.421.1043S} suggested some alternative explanations, like the separation between the WR stars and the regions emitting \ion{He}{II}, or SPs with higher temperatures than those expected by the current models.
While they discarded X-rays as a potential solution, the now-established relation between increased X-ray emission and the number of XRBs in galaxies with lower metallicity was not fully explored at the time.

The SFGs of the agnostic-to-X-rays sample examined in Section \ref{sec:All_HeII} extend from normal to very active galaxies as observed by their EW ($\rm 17 < EW_{H\beta}/$\AA$ < 411$).
The comparison between the SFGs' \ion{He}{II}/\ion{H}{$\beta$} ratio and $\rm EW_{H\beta}$ shows a lack of or even an anti-correlation.
This is against populations of WR stats being the dominant mechanism of \ion{He}{II} ionization since otherwise we should observe higher \ion{He}{II}/\ion{H}{$\beta$} ratios in galaxies with higher EWs and thus recent star-forming activity.
These sources range from extremely metal-poor to super-solar metallicity SFGs (7.67 < 12 + log(O/H) < 9.15; Figure \ref{fig:HeII_OIII_OII}).
The comparison does not show a correlation extending the full metallicity range of the sample. 
However, although the scatter is considerable, there is a weak positive correlation between metallicity and the \ion{He}{II}/\ion{H}{$\beta$} ratio at $\rm 12+log(O/H) \gtrsim 8.25$ which aligns with increased contribution from WR stars.
Therefore, while the excitation of \ion{He}{II} by WR stars cannot be ruled out, their influence demonstrates a lower degree of effect than that exerted by the XRB's X-ray emission.

\subsection{Possibility of X-ray obscured AGNs}

In this work, we do not fit separately the X-ray spectra, and our comparisons are based on the fluxes provided by the CSC.
The CSC does not fit the X-ray spectra of any source with models accounting for intrinsic absorption.
However, this could be the case for some of this work's Seyfert II AGNs, which were found to have hard X-ray spectra potentially attributed to intrinsic obscuration (Section \ref{sec:Hardness_obscuration}).
These sources also show higher gas-phase extinction (Figure \ref{fig:histogramms}).
We also do not correct the \ion{He}{II} emission for extinction.
Thus, this comparison involves the observed fluxes directly.
Further analysis accounting for X-ray obscuration and gas-phase extinction correction can provide a clearer picture of the X-ray and \ion{He}{II} luminosity relation.
Still, the compared parameters retain a tight correlation. 

\subsection{X-rays as the source of \texorpdfstring{\ion{He}{II}}{} excitation in AGNs and SFGs}

The behavior displayed by AGNs could be explained by assuming that the central BH is the dominating ionizing source in their host galaxies. 
The accretion disk of the central BH is capable of producing X-ray and hard UV emission \citep[e.g.,][]{1996ApJ...461...20H}. 
The spectral shape of this radiation and its flux depend on several parameters, such as the BH mass or accretion rate.
Assuming BHs of similar mass, sources with increased accretion rate would produce higher fluxes in both UV and X-ray emission \citep[e.g.,][]{2014ApJ...788...48S} capable of exciting \ion{He}{II} and elements that do not have such high excitation thresholds like hydrogen or oxygen.
Thus, a common source of ionizing radiation can interpret the observed correlation between \ion{He}{II}/\ion{H}{$\beta$}, O32, and X-ray emission in AGNs.

SFGs demonstrate a different behavior compared to AGNs, with a lack of correlation between \ion{He}{II}/\ion{H}{$\beta$} and O32 or $\rm EW_{H\beta}$, but a very tight correlation between O32 and $\rm EW_{H\beta}$ (Figure \ref{fig:HeII_OIII_OII}).
This behavior could be explained if we assume that there can be two main ionizing components in SFGs, the SP's UV light and the XRB's X-ray emission.
The dominating low-energy (lower than 54~eV required for the excitation of \ion{He}{II}) ionization source is mainly the young massive stars.
The young SPs dominate the low-energy ionizing radiation with higher flux compared to X-rays and that is why we observe the very tight correlation between O32 and the $\rm EW_{H\beta}$ in SFGs.

At the same time, in SFGs, the \ion{He}{II} emission remains unrelated to O32 and the $\rm EW_{H\beta}$ but is strongly correlated with the X-ray emission.
It is known that in SFGs the UV luminosity scales linearly with the \ion{H}{$\alpha$}, and subsequently the \ion{H}{$\beta$} luminosity, since hydrogen in the ISM is being excited by the same UV photons \citep[e.g.,][]{2002A&A...383..801B,2013Ap&SS.343..361P,2013ApJ...765...26S}.
The SFGs of our sample show a lack of or an anti-correlation between the \ion{He}{II}/\ion{H}{$\beta$} ratio and the $\rm EW_{H\beta}$ indicating that while the hydrogen emission-line and the UV luminosity strengthen they are not necessarily followed by the \ion{He}{II} emission.
As theorized in the past, this indicates that even abundant UV ionizing radiation from young SPs can not excite \ion{He}{II} if they are not in rare homogeneous O stars associations which can reach extremely high $T_{\rm eff}$ \citep[e.g.,][]{2013ApJ...766...91J,2015A&A...576A..83S}.
However, XRBs can abundantly produce high-energy radiation capable of ionizing \ion{He}{II}.

The integrated X-ray intensity of a population of high-mass XRBs is known to correlate with the SFR of the host galaxies \citep[e.g.,][]{2004MNRAS.347L..57G} with dependencies on the age of the SPs \citep[e.g.,][]{2013ApJ...776L..31F} or the metallicity \citep[e.g.,][]{2013ApJ...769...92P}. 
However, the presence of XRBs is rather stochastic \citep[e.g.,][]{2015A&A...579A..44D} with possibly elevated outputs in low-metallicity dwarf starbursts \citep[e.g.,][]{2022A&A...661A..16V} while the correlation between X-ray emission and the SFR shows significant scatter especially in lower masses and spatial scales \citep[e.g.,][]{2020MNRAS.494.5967K}.
The SFR of galaxies is traced by the strength of the emission of Balmer lines like \ion{H}{$\beta$}, and based on the results of this work the \ion{He}{II} emission is powered by the X-ray emission of the XRBs.
Thus, the observed stochasticity observed in the SFR-$L_{\rm X}$ relations could also partially explain the observed scatter of the \ion{He}{II}/\ion{H}{$\beta$} ratio found in SFGs. 

However, further examination is needed to fully understand the $L_{\ion{He}{II}}$--$L_{\rm X}$ correlation and its scatter. 
How does the \ion{He}{II} intensity depend on the amount of gas present and the state of the ISM?
Is the \ion{He}{II} emission following the X-ray emission at spatially resolved subgalactic scales, and to what spatial extent can the X-ray emission ionize Helium?
Do all galaxies with X-rays show \ion{He}{II} emission, and vice versa, do all galaxies with \ion{He}{II} emission emit X-rays?

For instance, \cite{2017MNRAS.472.2608S} reported one source with \ion{He}{II} emission but no X-ray equivalent.
However, based on this work's $L_{\ion{He}{II}}$--$L_{\rm X}$ correlation, the observed \ion{He}{II} luminosity $L_{\rm \ion{He}{II}} = 2.6 ~ 10^{37}~{\rm erg~s^{-1}}$ of the source SB~111 corresponds to $L_{\rm X} = 3.6 ~ 10^{38}~{\rm erg~s^{-1}}$ in the 0.5--1.2~keV energy range which would make the source marginally undetected for a limiting luminosity of $L_{\rm X} \simeq 3 ~ 10^{38}~{\rm erg~s^{-1}}$ in case it has a soft X-ray spectrum as most of our sample's SFGs.
Similarly, \cite{2018MNRAS.480.1081K} which examined the metal-poor galaxy SBS 0335-052E, suggested that the observed \ion{He}{II} intensity could not be fully explained by the relatively low X-ray luminosity.
However, based on this work's $L_{\ion{He}{II}}$--$L_{\rm X}$ correlation and accounting for the scatter and the lower energy range, its 0.5--5~keV $L_{\rm X} = 4 ~ 10^{39}~{\rm erg~s^{-1}}$ is in agreement with the source's observed $L_{\rm He~II} = 2.16 ~ 10^{38}~{\rm erg~s^{-1}}$.

The study of \cite{2004MNRAS.351L..83K} performed a spatially resolved analysis of a bright X-ray source located at the outskirts of the dwarf irregular galaxy Holmberg II. 
The source is considered to be a ULX with $L_{\rm X} = 5 ~ 10^{39}~ {\rm erg~s^{-1}}$.
This source was found to be surrounded by a \ion{H}{II} region emitting \ion{He}{II} with $L_{\rm \ion{He}{II}} = 2.7 ~ 10^{36}~ {\rm erg~s^{-1}}$.
Moreover, the location of the \ion{He}{II} emission in Holmberg II follows spatially the X-ray source.
The observed luminosities agree with this work's $L_{\ion{He}{II}}$--$L_{\rm X}$ relation considering the scatter and uncertainties.
In this case, the observed $L_{\rm \ion{He}{II}}$ is weaker and not stronger with respect to the expected by the $L_{\rm \ion{He}{II}}$--$L_{\rm X}$ relation as in the cases of SB~111 and SBS 0335-052E.

The results of this work are also similar to that of \cite{2004A&A...414..825S} which have reported a linear correlation between the X-ray luminosity and that of another high ionization but mid-IR emission line, the [\ion{O}{IV}] at $\lambda = 26 \mu$m.
Similarly to \ion{He}{II}, the excitation of $\rm [\ion{O}{IV}]$ requires photons with energies higher than 55~eV.
The $L_{\rm X}$--$ L_{\rm [\ion{O}{IV}]}$ comparison reported by \cite{2004A&A...414..825S} involved starbursts, LINERs, and AGN galaxies as classified by the morphology and distribution of their X-ray sources showing a correlation transcending their classes. 

Overall, the results of this work are aligned with previous studies proposing the XRBs' X-ray emission as the main source of \ion{He}{II} ionization in low metallicity galaxies \citep[e.g.,][]{2019A&A...622L..10S,2019A&A...627A..63O,2021A&A...656A.127S,2022A&A...661A..67O,2022ApJ...930..135L}.
Using binary stellar evolution simulations \citep{2013ApJ...764...41F,2013ApJ...776L..31F}, \cite{2019A&A...622L..10S} modeled the expected relationship between the \ion{He}{II}/\ion{H}{$\beta$} ratio and $\rm EW_{H\beta}$, incorporating the contributions of XRBs associated with the parent SPs.
The SFGs analyzed in this study (Figure \ref{fig:HeII_OIII_OII}) align well with these models, exhibiting \ion{He}{II}/\ion{H}{$\beta$} ratios significantly higher than those predicted by the \cite{2018MNRAS.477..904X} binary evolution models \citep[BPASS;][]{2017PASA...34...58E}, which do not account for XRB contributions.
Additionally, our analysis demonstrates that the correlation between X-ray and \ion{He}{II} luminosities extends to SFGs with higher, near-solar metallicities. 
These galaxies show \ion{He}{II}/\ion{H}{$\beta$} ratios and $\rm EW_{H\beta}$ values consistent with the higher-metallicity models presented in \cite{2019A&A...622L..10S}.

Moreover, this work shows that the $L_{\ion{He}{II}}$--$L_{\rm X}$ correlation holds regardless of the class of the galaxies and extends to AGNs which show on average higher luminosities than SFGs.
The separate fits for the AGNs and SFGs overlap suggesting a common \ion{He}{II} excitation mechanism regardless if the X-ray source is a central super-massive BH or XRBs scattered throughout the host galaxies.

\section{Conclusions}
\label{sec:conclusions}

Summarizing the results of this work, we crossmatched the CSC with the \citetalias{2012MNRAS.421.1043S} and MPA-JHU catalogs to examine the relation of \ion{He}{II} and X-ray emission in galaxies.
We analyzed 165 galaxies well detected in X-rays by Chandra and \ion{He}{II} by SDSS, and 2399 galaxies with \ion{He}{II} emission matched with the MPA-JHU catalog regardless if they have X-ray observations.

Regarding the X-ray and \ion{He}{II} emitting galaxies:

\begin{itemize}
    \item BPT SFGs classified as AGNs by \citetalias{2012MNRAS.421.1043S} show high metallicities, indicating evolved and passive host galaxies contradicting their strong \ion{He}{II}/\ion{H}{$\beta$} ratio.
    Additionally, these sources show high-velocity dispersions and excess X-ray emission compared to what is expected by their SFR. 
    Most of them also show broadened Balmer lines compared to the forbidden ones and are considered NLS1s in this work.

    \item On average, our sample's SFGs show a softer X-ray spectrum compared to NLS1s while the Seyfert II AGNs show the hardest.
    Similarly, SFGs and NLS1s show on average lower gas-phase extinction than Seyfert II AGNs.
    This suggests that many of our sample's Seyfert II AGNs may display an obscured X-ray spectrum.
    
    \item The $L_{\rm X}$/SFR of SFGs is linearly correlated with the \ion{He}{II}/\ion{H}{$\beta$} ratio. 
    This relation holds for all X-ray energy bands provided by CSC and it shows a lower scatter than the $L_{\rm X}$/SFR--metallicity relation. 
    Furthermore, SFGs with higher sSFR tend to fall below the best-fit $L_{\rm X}$/SFR--\ion{He}{II}/\ion{H}{$\beta$} relation, while those with slightly lower sSFR are above, suggesting a dependence on recent star-forming activity and stellar population age.

    \item There is a unified tight correlation between the observed X-ray and \ion{He}{II} luminosity that holds for both SFGs and AGNs.
    While the former show lower luminosities on average compared to the latter, the linear regression fits overlap indicating a similar \ion{He}{II} excitation mechanism.
    This correlation remains similarly tight for about seven orders of magnitude of X-ray and \ion{He}{II} luminosities.
    Based on MCMC fitting, we provide linear relations separately for the SFGs, AGNs, the total sample, and the different X-ray energy bands.

\end{itemize}
Regarding the \citetalias{2012MNRAS.421.1043S} \ion{He}{II} galaxies regardless if they were observed in X-rays:

\begin{itemize}

    \item AGNs and SFGs show significant differences in the comparisons between \ion{He}{II}/\ion{H}{$\beta$}, O32 ratio, and the $\rm EW_{H\beta}$.
    The O32 ratio of purely SFGs is tightly correlated with their $\rm EW_{H\beta}$. 
    This correlation indicates that the SFGs star-forming activity dominates the ionization of lower energy species.
    
    \item AGNs show a higher O32 ratio but are less actively star-forming as indicated by their lower $\rm EW_{H\beta}$. 
    The O32 ratio compared to the $\rm EW_{H\beta}$ of AGNs is significantly higher than that of SFGs showing a large scatter and no apparent correlation. 
    This could be attributed to the effect of the AGN's ionizing radiation on the ISM of the host galaxies.
    We derive a line above which only AGNs of our sample are located.
        
    \item The O32 ratio is correlated with the \ion{He}{II}/\ion{H}{$\beta$} ratio in AGNs. 
    This indicates that the AGN's radiation is the dominating ionization source over the host galaxies' ISM for low and higher ionization species. 
    
    \item The O32 ratio is not correlated with the \ion{He}{II}/\ion{H}{$\beta$} ratio in SFGs. 
    This, along with the tight correlation between the O32 ratio and $\rm EW_{H\beta}$, suggests that while the lower-energy ionization is attributed to star-forming activity, high-energy species like \ion{He}{II} require a different source of excitation.
    The tight correlation between X-ray and \ion{He}{II} emission favors that the latter is excited by the higher energy but lower overall flux X-ray photons produced by X-ray binary populations.
    
\end{itemize}

This work suggests that X-ray emission is the main source of high-energy \ion{He}{II} ionization in galaxies, regardless of whether the ionizing source is an AGN or a population of XRBs.
Based on archival homogenized data, we provide linear fits that quantify the correlation between the X-ray and \ion{He}{II} luminosities.
However, further detailed analysis is required to examine the effects of X-ray obscuration, gas-phase extinction, and universality of the correlation.
Moreover, to fully constrain the underlying physical processes, further comparisons with modeling of the photoionization of the ISM and the radiation transfer concerning X-ray and UV radiation should be performed.

\begin{acknowledgements}

The authors thank the anonymous referee for providing comments and suggestions that improved this work.
The Czech Science Foundation project No. 22-22643S supported this work.
This research has made use of: 
(a) software provided by the CSC in the application packages DS9;
(b) the NASA/IPAC Extragalactic Database (NED), which is operated by the Jet Propulsion Laboratory (JPL), California Institute of Technology, under contract with NASA; 
(c) the NASA/IPAC Infrared Science Archive (IRSA), which is funded by NASA and operated by the California Institute of Technology; 
Funding for the SDSS and SDSS-II has been provided by the Alfred P. Sloan Foundation, the Participating Institutions, the National Science Foundation, the U.S. Department of Energy, the National Aeronautics and Space Administration, the Japanese Monbukagakusho, the Max Planck Society, and the Higher Education Funding Council for England. The SDSS Web Site is http://www.sdss.org/.

The SDSS is managed by the Astrophysical Research Consortium for the Participating Institutions. The Participating Institutions are the American Museum of Natural History, Astrophysical Institute Potsdam, University of Basel, University of Cambridge, Case Western Reserve University, University of Chicago, Drexel University, Fermilab, the Institute for Advanced Study, the Japan Participation Group, Johns Hopkins University, the Joint Institute for Nuclear Astrophysics, the Kavli Institute for Particle Astrophysics and Cosmology, the Korean Scientist Group, the Chinese Academy of Sciences (LAMOST), Los Alamos National Laboratory, the Max-Planck-Institute for Astronomy (MPIA), the Max-Planck-Institute for Astrophysics (MPA), New Mexico State University, Ohio State University, University of Pittsburgh, University of Portsmouth, Princeton University, the United States Naval Observatory, and the University of Washington.

This research made use of the Digitized Sky Surveys (DSS).
The Digitized Sky Surveys were produced at the Space Telescope Science Institute under U.S. Government grant NAG W-2166. The images of these surveys are based on photographic data obtained using the Oschin Schmidt Telescope on Palomar Mountain and the UK Schmidt Telescope. The plates were processed into the present compressed digital form with the permission of these institutions.
The National Geographic Society - Palomar Observatory Sky Atlas (POSS-I) was made by the California Institute of Technology with grants from the National Geographic Society.
The Second Palomar Observatory Sky Survey (POSS-II) was made by the California Institute of Technology with funds from the National Science Foundation, the National Geographic Society, the Sloan Foundation, the Samuel Oschin Foundation, and the Eastman Kodak Corporation.
The Oschin Schmidt Telescope is operated by the California Institute of Technology and Palomar Observatory.
The UK Schmidt Telescope was operated by the Royal Observatory Edinburgh, with funding from the UK Science and Engineering Research Council (later the UK Particle Physics and Astronomy Research Council), until 1988 June, and thereafter by the Anglo-Australian Observatory. The blue plates of the southern Sky Atlas and its Equatorial Extension (together known as the SERC-J), as well as the Equatorial Red (ER), and the Second Epoch [red] Survey (SES) were all taken with the UK Schmidt.
All data are subject to the copyright given in the copyright summary. Copyright information specific to individual plates is provided in the downloaded FITS headers.
Supplemental funding for sky-survey work at the ST ScI is provided by the European Southern Observatory. 
This research has made use of TOPCAT \citep{2005ASPC..347...29T}.

\end{acknowledgements}

\bibliographystyle{aa} 
\bibliography{citations.bib} 

\begin{appendix}
\section{Optical spectra of the X-ray BPT SFGs but SB12 AGNs.}
\label{sec:Appendix_A}

This appendix presents rest-frame optical spectra near the \ion{H}{$\alpha$} and \ion{H}{$\beta$} lines of this work's 30 X-ray-detected sources which were classified as SFGs by BPT but as AGNs by the SB12 diagram (Section \ref{sec:classification}). 
Based on visual inspection, the spectra of 29/30 sources show broadened Balmer lines demonstrating contribution from an AGN's BLR.

\begin{figure}[ht!]
    \centering
    \includegraphics[width=0.48\textwidth]{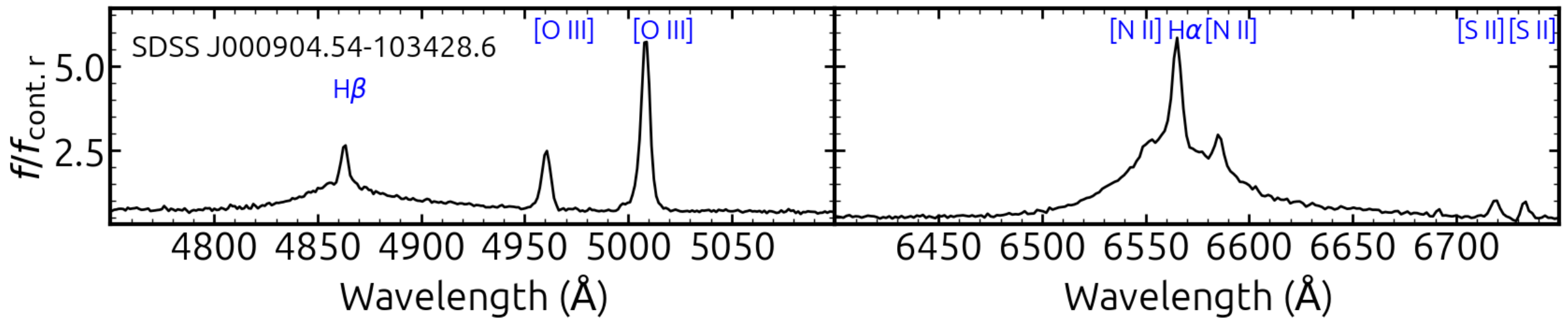}
    \includegraphics[width=0.48\textwidth]{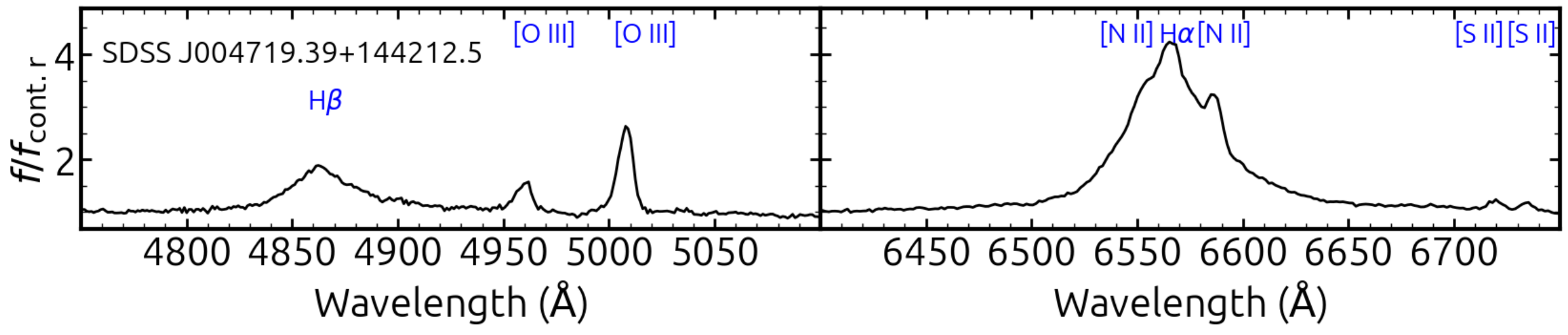}
    \includegraphics[width=0.48\textwidth]{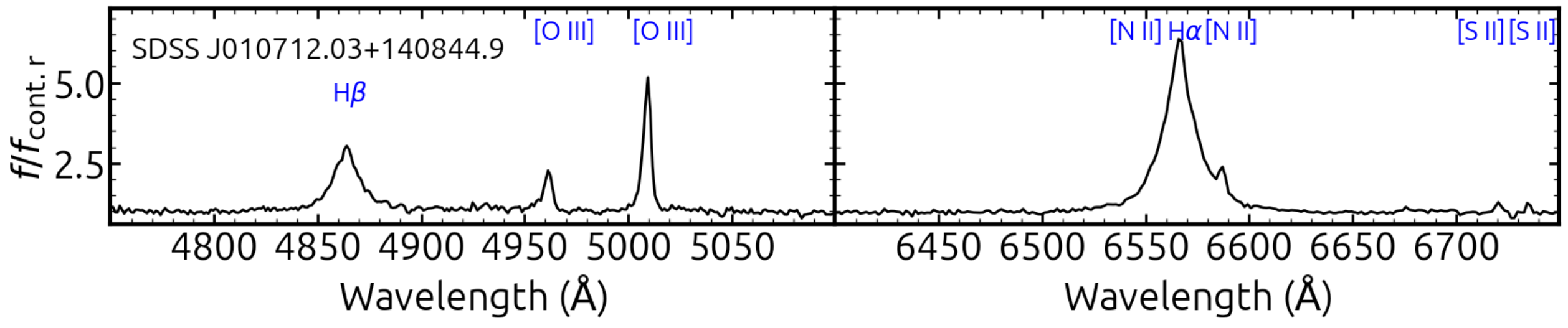}
    \includegraphics[width=0.48\textwidth]{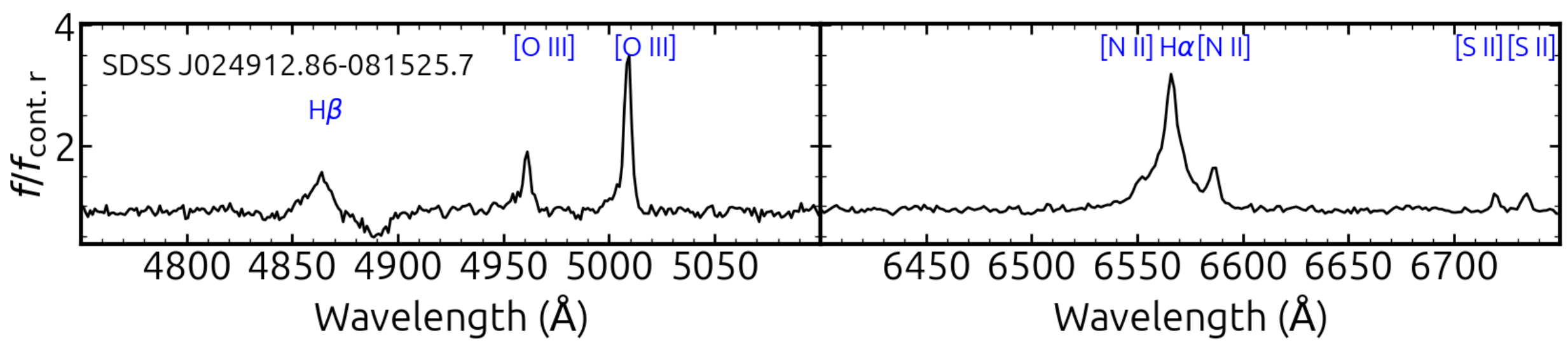}
    \includegraphics[width=0.48\textwidth]{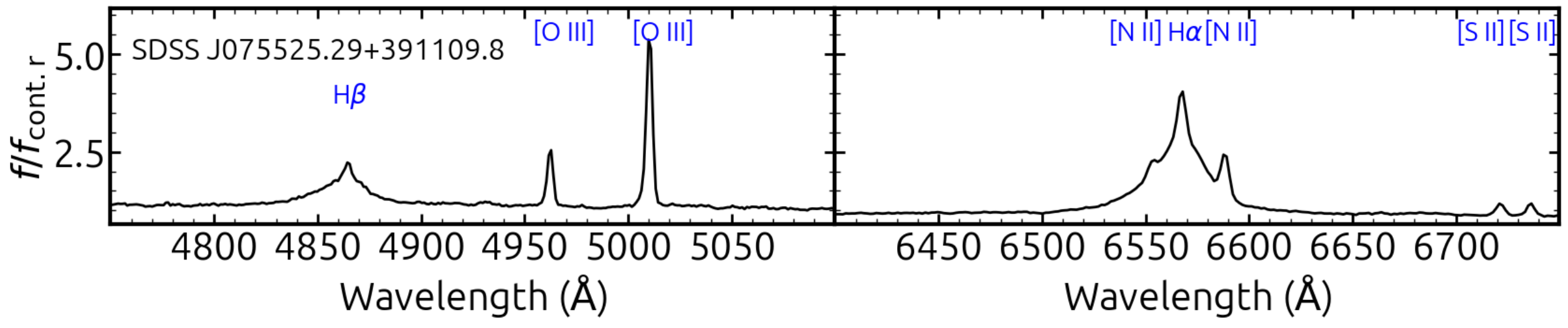}
    \includegraphics[width=0.48\textwidth]{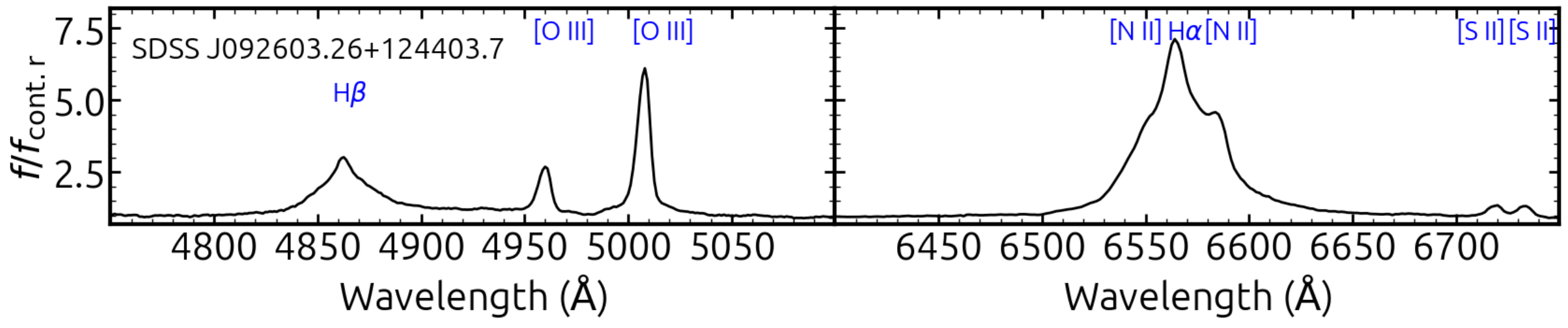}
    \includegraphics[width=0.48\textwidth]{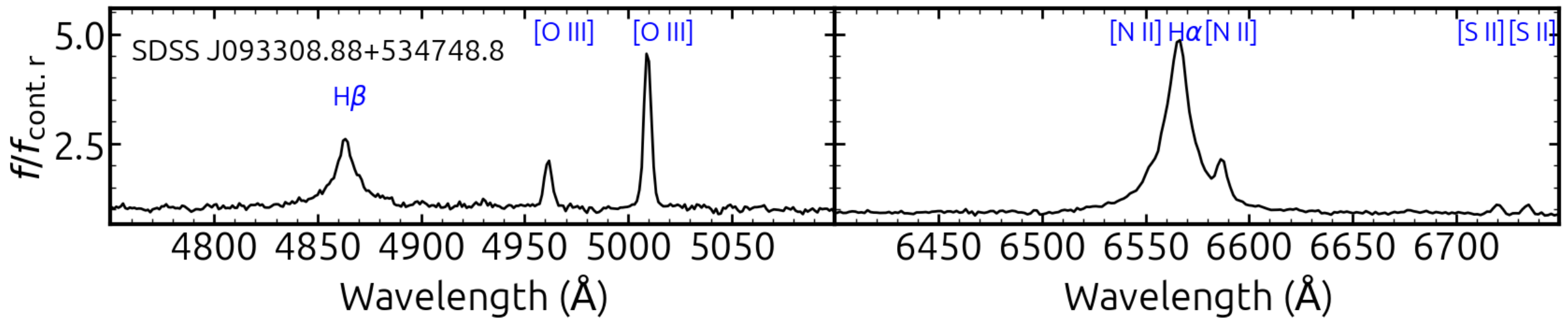}
    \includegraphics[width=0.48\textwidth]{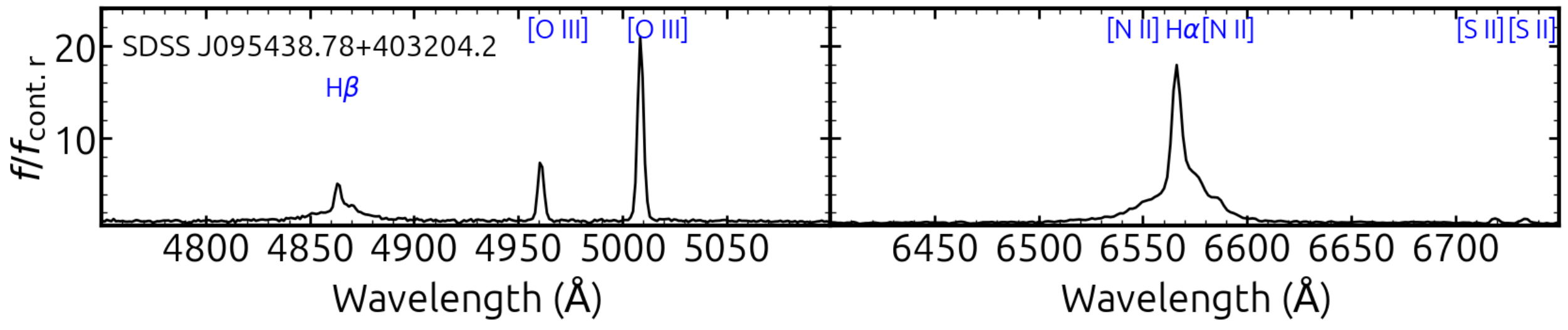}
    \includegraphics[width=0.48\textwidth]{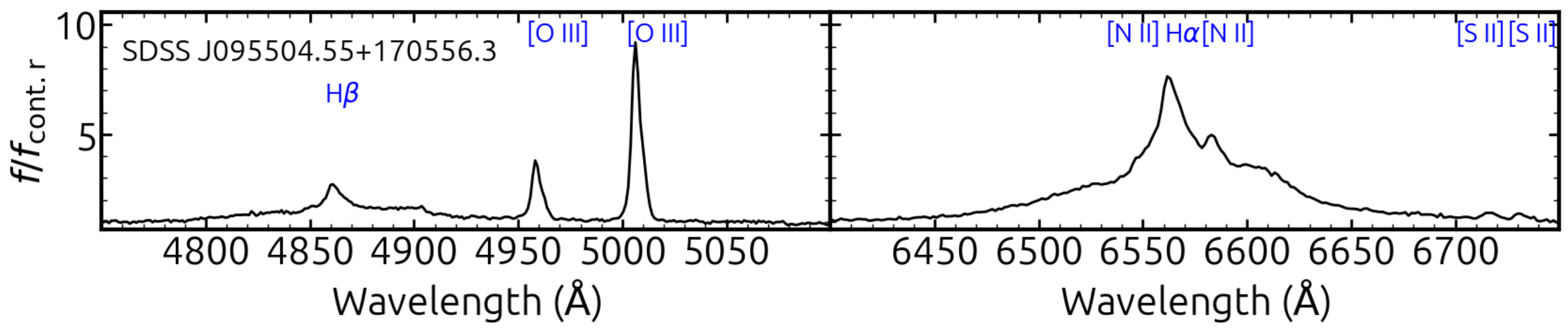}
    \includegraphics[width=0.48\textwidth]{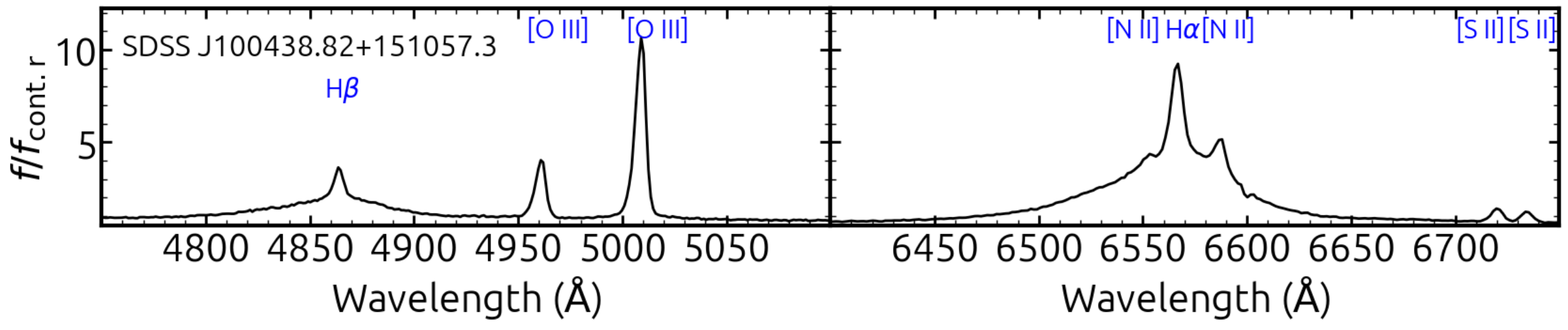}
    \caption{Optical spectra near the \ion{H}{$\beta$} (left plot) and \ion{H}{$\alpha$} (right plot) emission lines of the X-ray and \ion{He}{II} detected BPT SFGs but SB12 AGNs (Section \ref{sec:classification}). }
    \label{fig:specs_broad_1}
\end{figure}

\begin{figure}
    \includegraphics[width=0.48\textwidth]{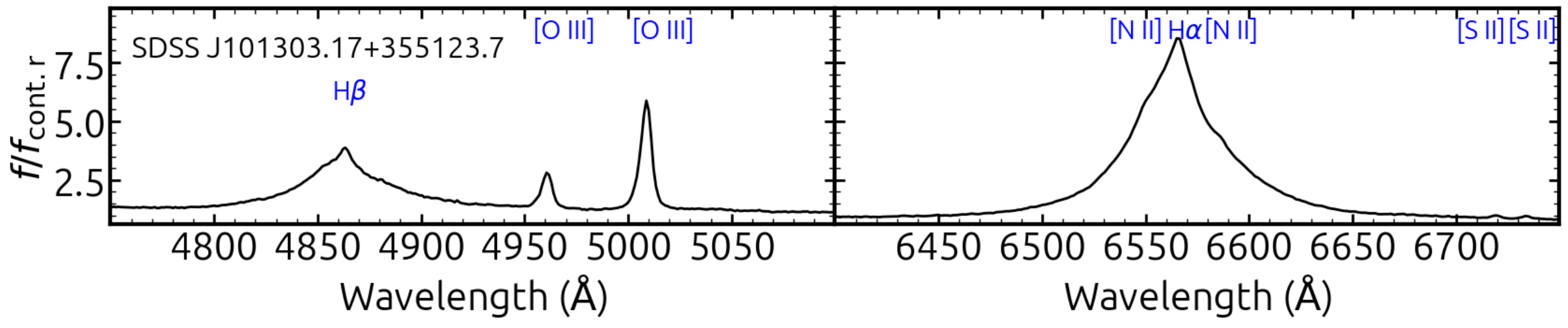}
    \includegraphics[width=0.48\textwidth]{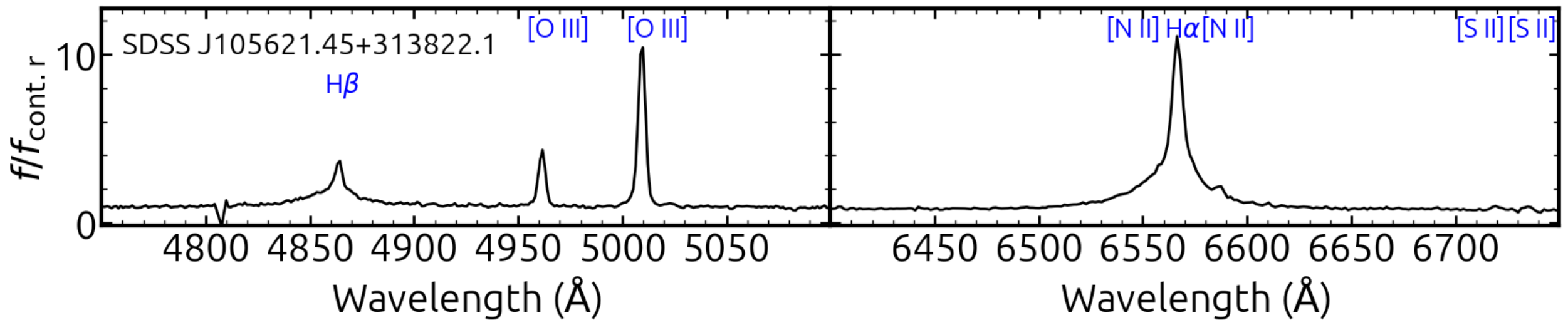}
    \includegraphics[width=0.48\textwidth]{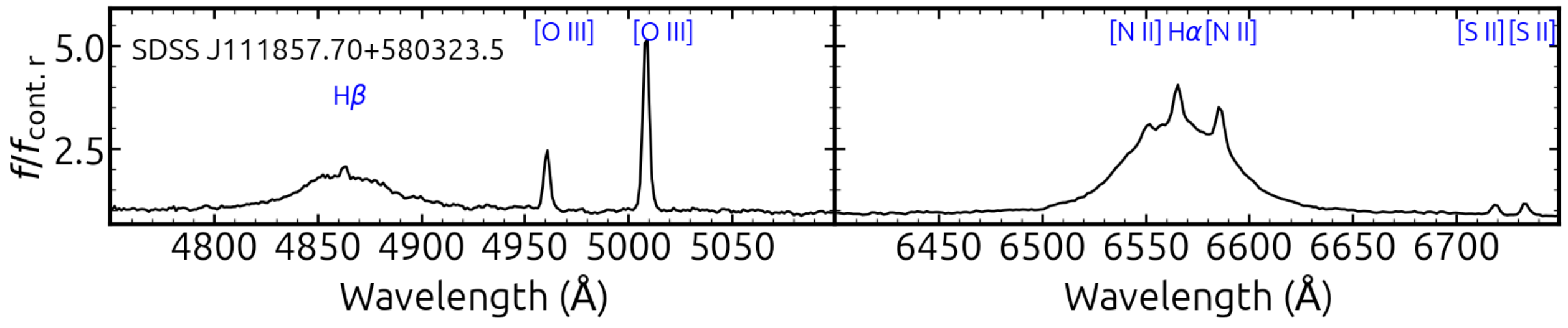}
    \includegraphics[width=0.48\textwidth]{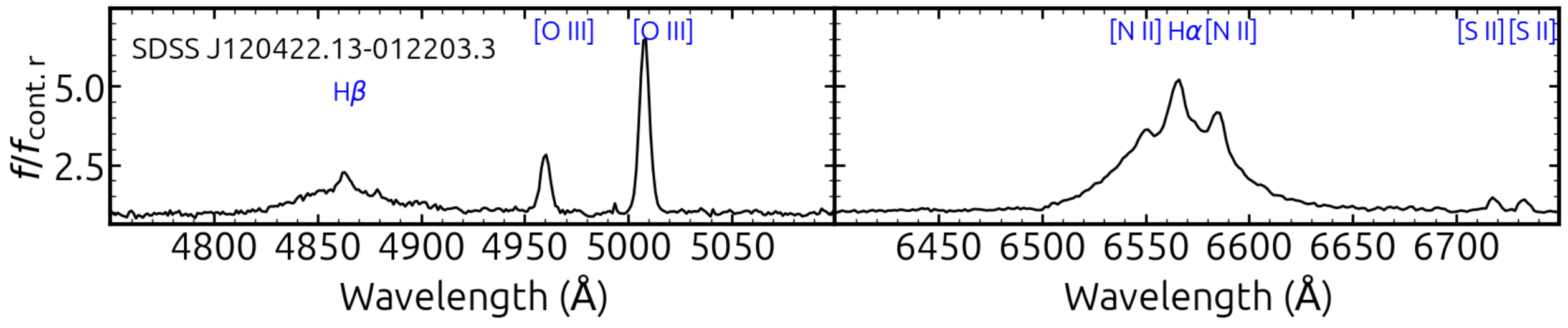}
    \includegraphics[width=0.48\textwidth]{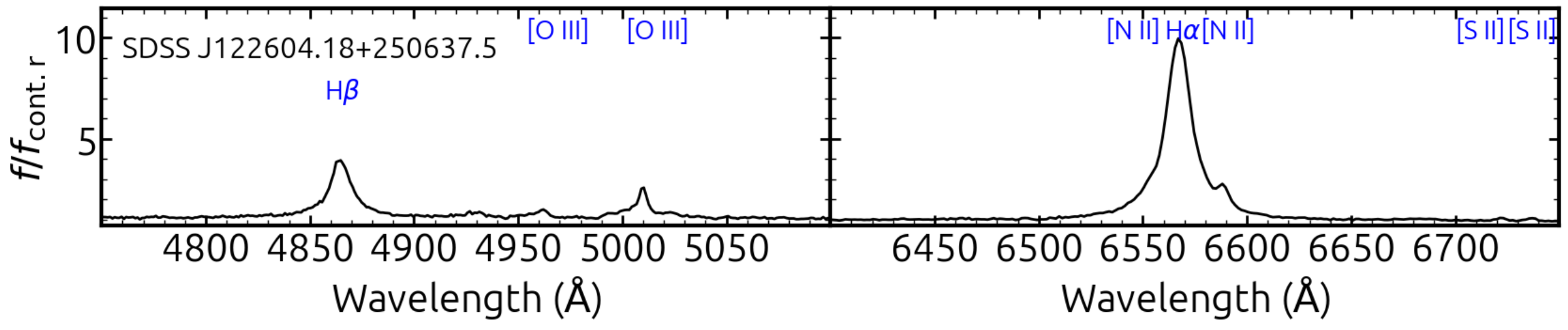}
    \includegraphics[width=0.48\textwidth]{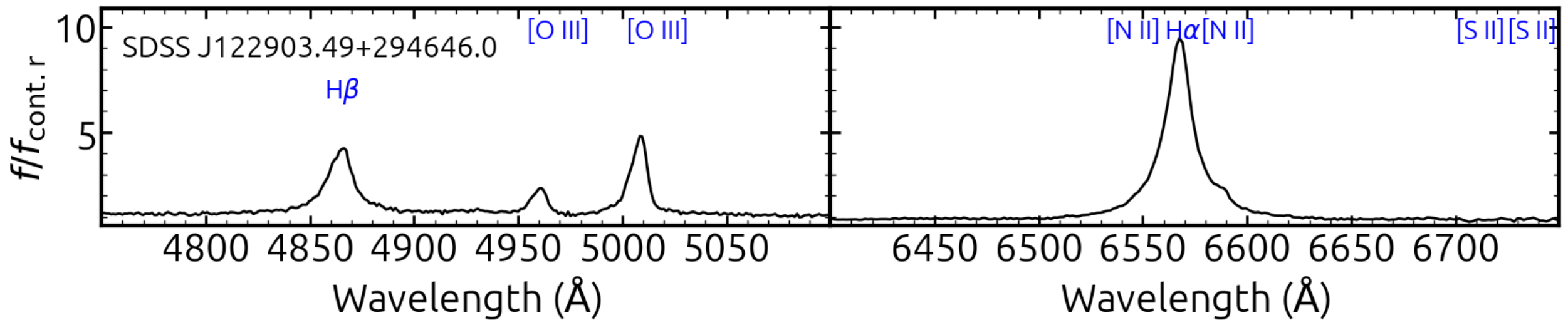}
    \includegraphics[width=0.48\textwidth]{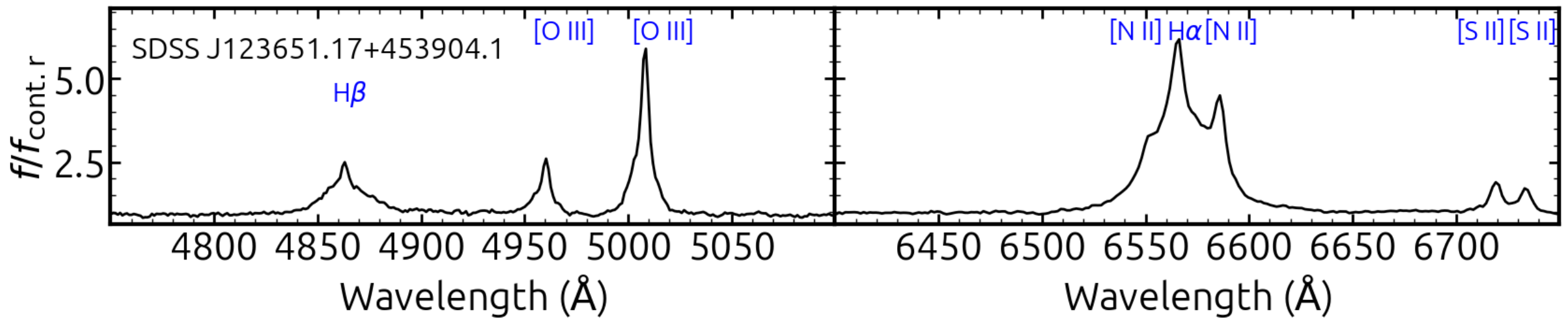}
    \includegraphics[width=0.48\textwidth]{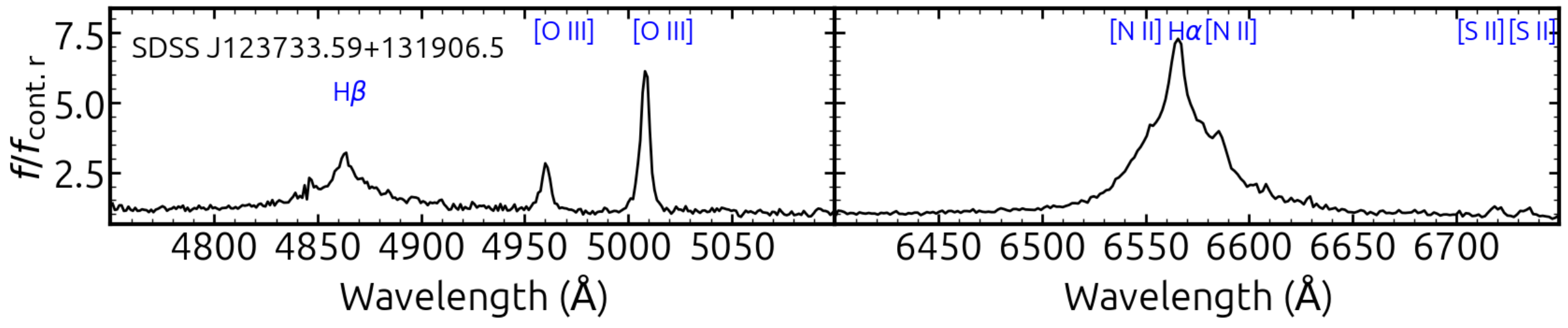}
    \includegraphics[width=0.48\textwidth]{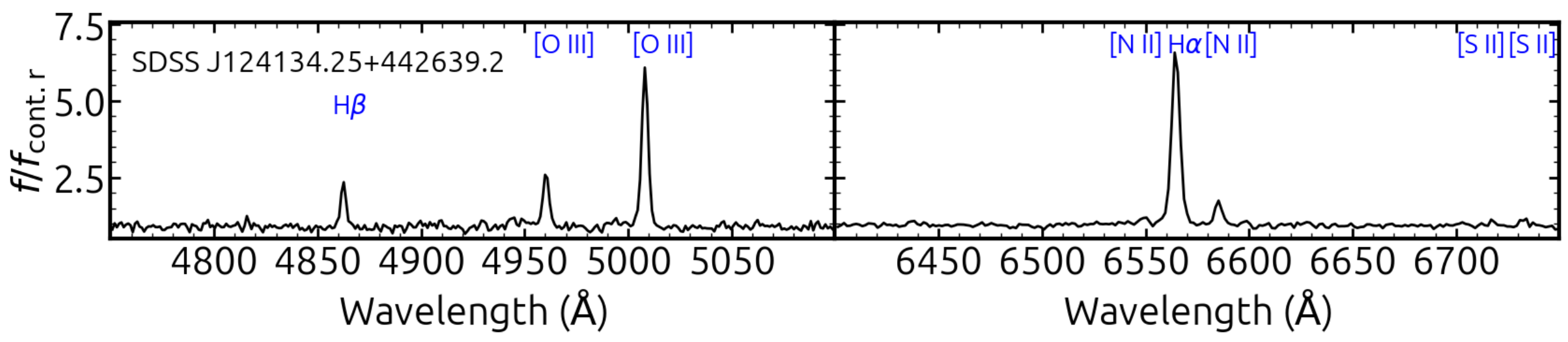}
    \includegraphics[width=0.48\textwidth]{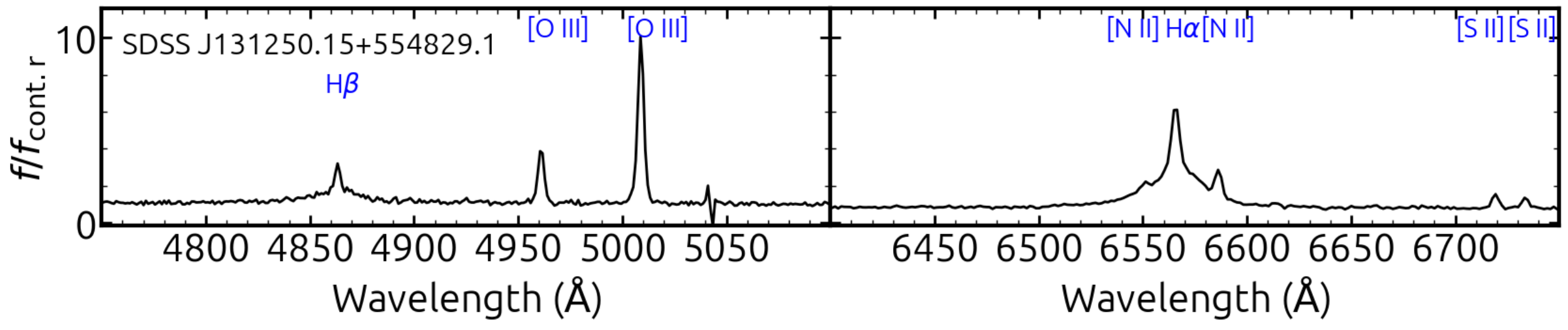}
    \includegraphics[width=0.48\textwidth]{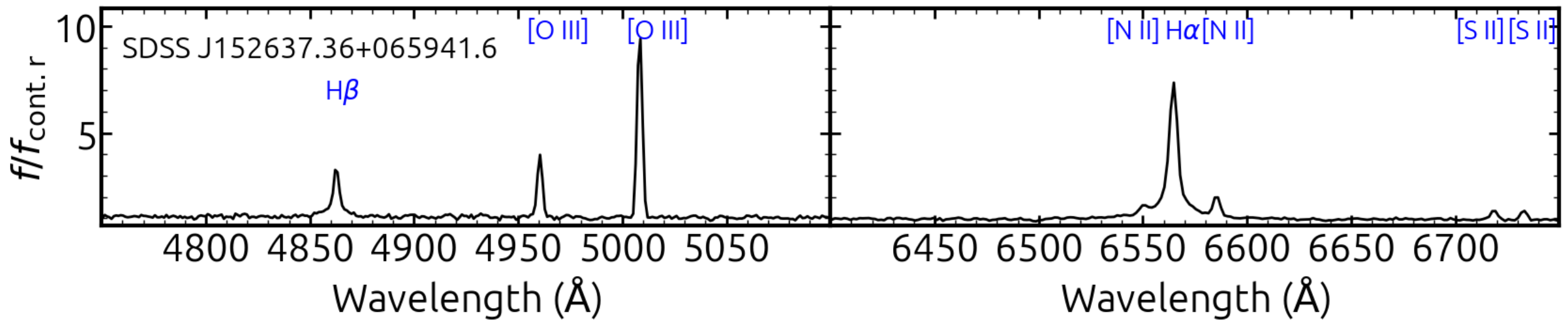}
    \includegraphics[width=0.48\textwidth]{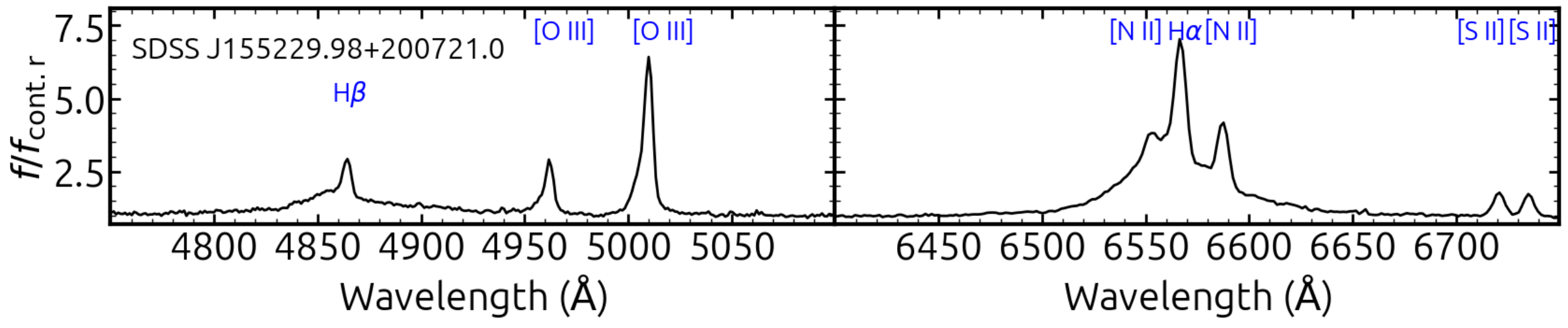}
    \caption{Continuation of Figure \ref{fig:specs_broad_1}}
    \label{fig:specs_broad_2}
\end{figure}

\begin{figure} 
    \includegraphics[width=0.48\textwidth]{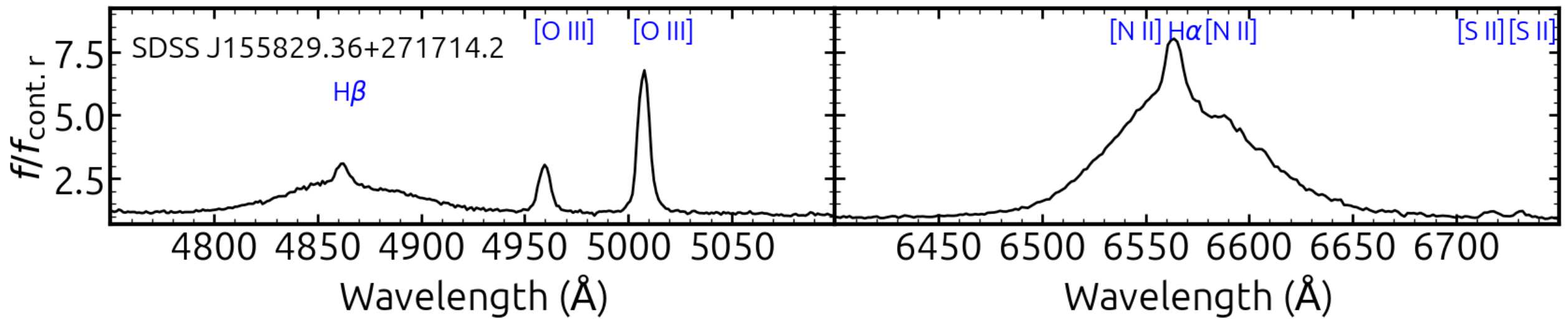}
    \includegraphics[width=0.48\textwidth]{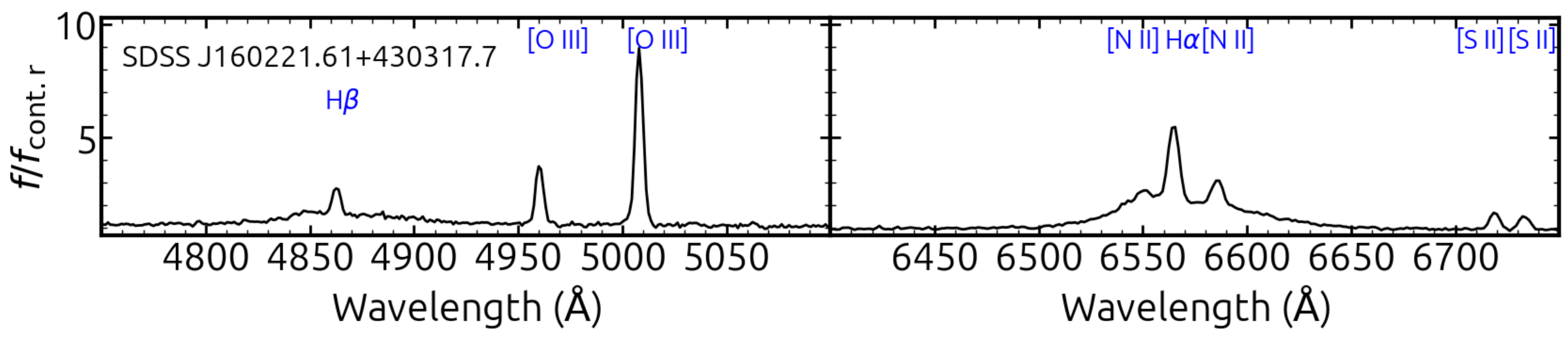}
    \includegraphics[width=0.48\textwidth]{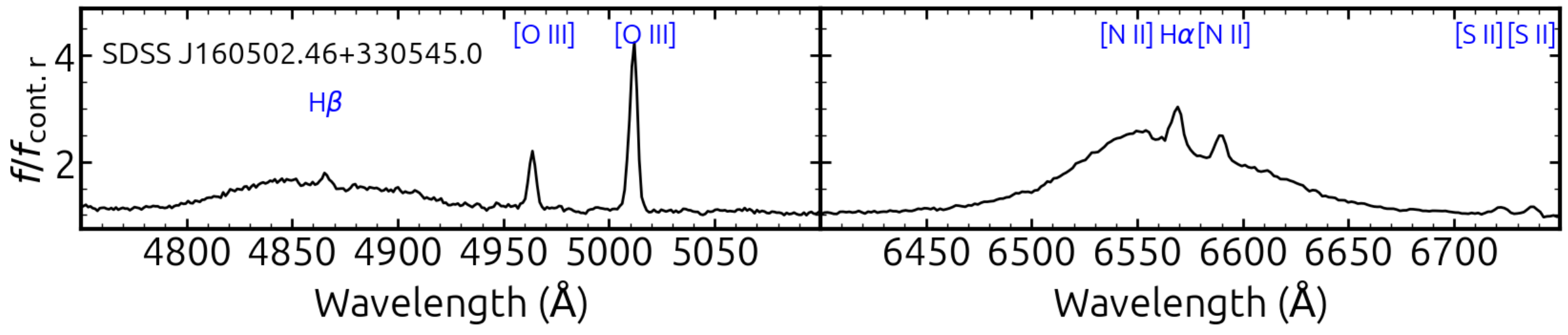}
    \includegraphics[width=0.48\textwidth]{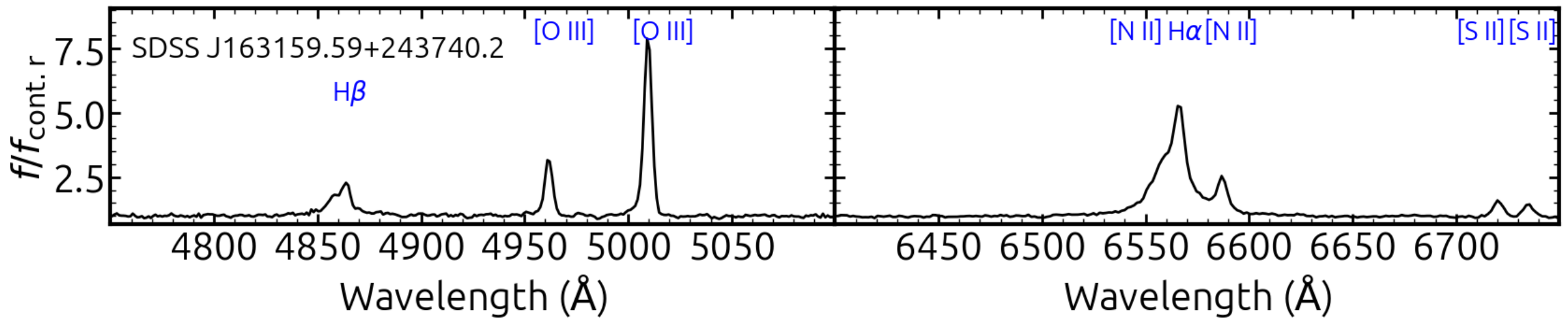}
    \includegraphics[width=0.48\textwidth]{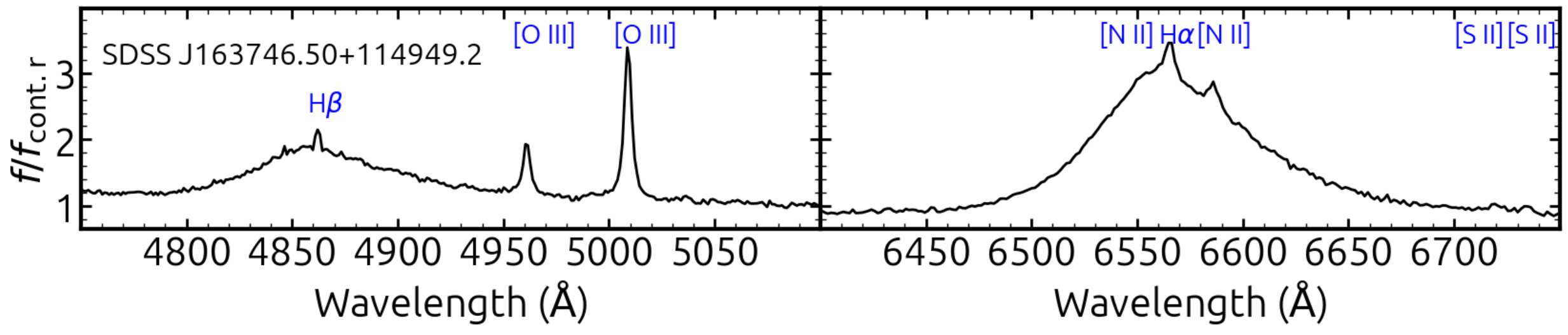}
    \includegraphics[width=0.48\textwidth]{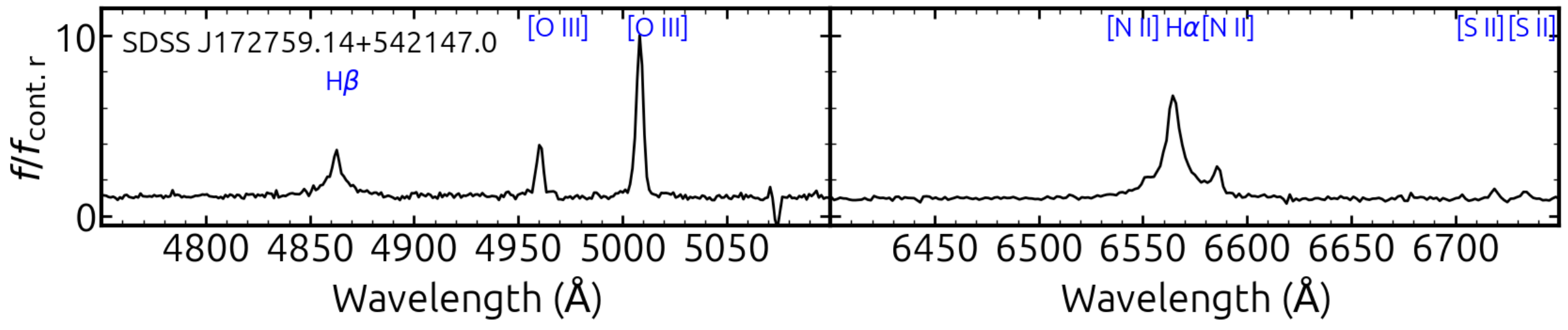}
    \includegraphics[width=0.48\textwidth]{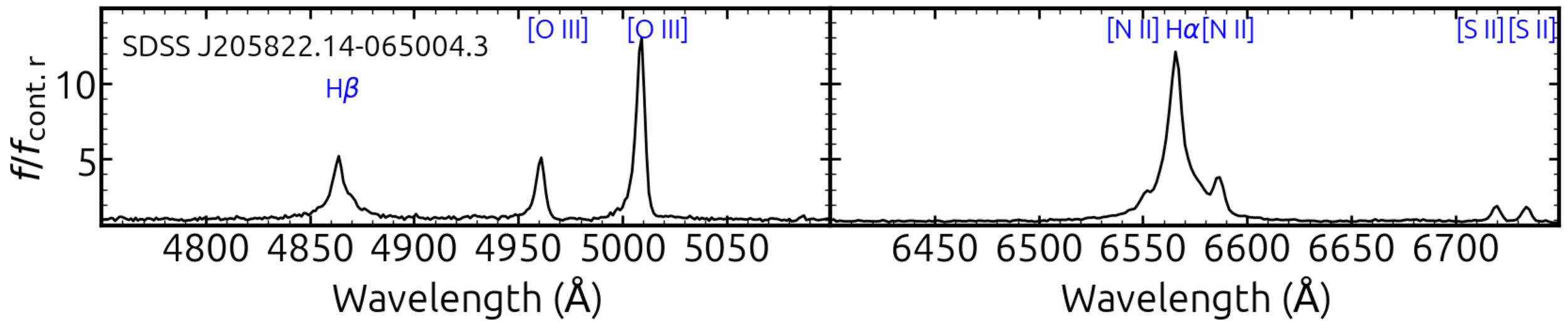}
    \includegraphics[width=0.48\textwidth]{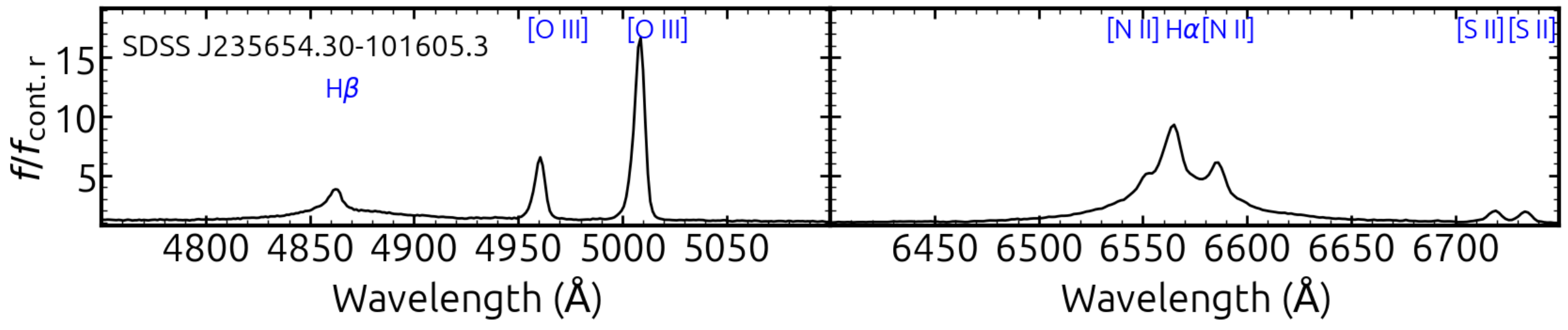}
    \caption{Continuation of Figure \ref{fig:specs_broad_2}}
    \label{fig:specs_broad_3}
\end{figure}

\end{appendix}

\end{document}